# Phase retrieval methods applied to coherent imaging

Tatiana Latychevskaia

Physics Institute, University of Zurich, Winterthurerstrasse 190, 8057 Zurich, Switzerland
Paul Scherrer Institute, Forschungsstrasse 111, 5232 Villigen, Switzerland
tatiana@physik.uzh.ch

# Contents

# 1. Introduction to the imaging of a non-crystalline object and the phase problem

Currently, the structures of biomolecules are typically obtained using X-ray crystallography, cryo-electron microscopy and nuclear magnetic resonance (NMR) techniques, which has resulted in an impressive database of molecular structures [1]. However, the structures obtained with these techniques are all generated as a result of averaging over many molecules, which involves averaging over fine conformational details. The goal of modern imaging techniques is to visualise an individual molecule at atomic resolution. The direct visualisation of an individual molecule at angstrom resolution could be achieved by using short-wavelength electron or X-ray waves. X-ray free-electron laser (XFEL) facilities in many countries around the world are being developed with the aim of providing such a tool for the visualisation of single biomolecules at atomic resolution.

Imaging without lenses is preferred, as this avoids aberrations and achieves the highest possible resolution. The principle of lensless imaging of a sample, for example an individual molecule, is as follows: when a coherent wave is scattered by a molecule, it carries both amplitude and phase information imposed by the scattering events. The phase distribution is especially important, since it contains information about the position of the atoms constituting the molecule. Detectors are not sensitive to phase information, however; they simply record the intensity, which is the square of the wave amplitude. Thus, to reconstruct the molecular structure, the phase of the complex-valued scattered wave must be recovered, and this constitutes the so-called *phase problem*. In essence, lensless imaging consists of the experimental acquisition of an interference pattern, followed by numerical phase retrieval of the molecular structures. An overview of the most popular lensless imaging schemes, that is, with no lenses between the sample and detector, is given in Fig. 1. Coherent diffractive imaging (CDI) and holography are the two basic types of coherent imaging technique, and the other techniques are derivatives of these approaches. Both of the imaging schemes shown in Fig. 1 are discussed in more detail in Section 2. In all of these imaging schemes, the experimentally recorded images are digitally sampled with $N \times N$ pixels and are subject to numerical reconstruction.

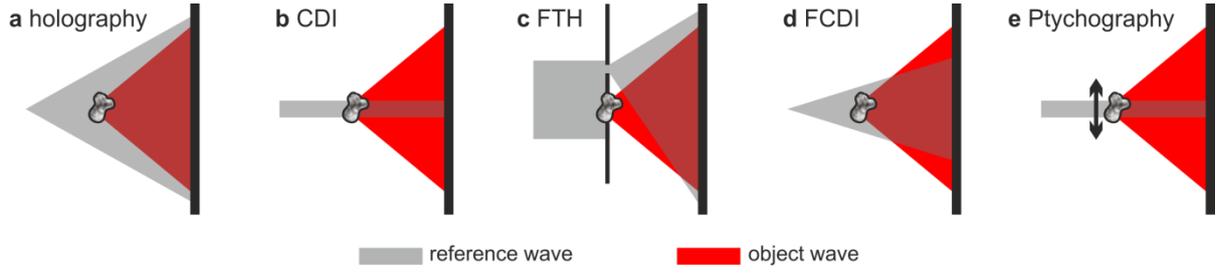

reference wave | object wave

Fig. 1. Experimental schemes for lensless imaging. (a) Gabor in-line holography. (b) Coherent diffraction imaging (CDI). (c) Fourier transform holography (FTH). (d) Fresnel coherent diffraction imaging (FCDI). (e) Ptychography.

**1.1 Coherence**

All of the imaging techniques shown in Fig. 1 require a coherent source of radiation. Coherence is a measure that characterises the stability of the phase difference between two interfering waves; the degree of coherence is measured by the visibility (contrast) of the interference pattern created [2, 3].

**Spatial (transverse) coherence** describes the correlation between the phases of the waves measured at different points in space. According to the van Cittert-Zernike theorem [4, 5], the complex coherence factor is given by the Fourier transform of the intensity distribution of the source [6]; that is, the spatial coherence is defined by the size of the source. For a source with an intensity distribution described by the Gaussian function $s(\xi,\eta) = \exp\left(-\frac{\xi^2 + \eta^2}{2\sigma^2}\right)$, where $(\xi,\eta)$ are the coordinates in the source plane and $\sigma$ is the standard deviation, the spatial coherence length at a distance $L$ from the source is given by [7]:

$$l_c^{\text{Spatial}} = \frac{\lambda L}{2\pi\sigma}, \tag{1}$$

where $\lambda$ is the wavelength. It follows from Eq. (1) that the smaller the source, the better the spatial coherence, and waves originating from an ideal point-like source are spatially infinitely coherent. Physical sources have a finite size, and thus have limited coherence, although this is sufficient for most imaging experiments. For example, for low-energy electrons of energy 250 eV (wavelength $\lambda$ = 0.078 nm), and source size $\sigma$ = 0.1 nm, the spatial coherence length is about 120 nm at a distance of 1 μm from the source, which is sufficient to image a macromolecule a few tens of nanometres in size, placed at about 1 μm in front of the electron source.

For a collimated beam conventionally employed in X-ray coherent imaging, the spatial coherence is defined by the divergence of the beam, $\Delta\vartheta$ [8]:

$$l_c^{\text{Spatial}} = \frac{\lambda}{2\Delta\vartheta}. \tag{2}$$

**Temporal coherence** is a measure of how monochromatic a source is. The temporal coherence length $l_c^{\text{Temporal}}$ of a wave with wavelength spread $\lambda \pm \Delta\lambda$ is proportional to

$$l_c^{\text{Temporal}} = \frac{\lambda^2}{\Delta\lambda}. \tag{3}$$

For photons, $\lambda = \frac{hc}{E}$, where $h$ is the Planck constant and $c$ is the speed of light. Thus, for a photon source with an energy spread $E \pm \Delta E$, the temporal coherence is given by:

$$l_c^{\text{Temporal}} = \frac{E}{\Delta E}\lambda. \tag{4}$$

For non-relativistic electrons, their wavelength is given by $\lambda = \frac{h}{\sqrt{2meU}}$, where $m$ is the electron's mass and $eU$ is its energy. The temporal coherence length is then calculated as:

$$l_c^{\text{Temporal}} = 2\frac{U}{\Delta U}\lambda. \tag{5}$$

As an example, Eq. (5) gives $l_c^{\text{Temporal}} \approx 390\,\text{nm}$ for low-energy electrons with $eU$ = 250 ± 0.1 eV.

## 1.2 Resolution

The achievable resolution is the key parameter of any imaging technique. In digital Gabor in-line holography, the resolution is given by [9, 10]:

$$R_{\text{Holography}} = \frac{\lambda Z}{N\Delta_H} = \frac{\lambda Z}{S_H}, \tag{6}$$

where $Z$ is the distance between the sample and the detector, $\Delta_H$ is the pixel size in the hologram plane, and $S_H = N\Delta_H$ is the side length of the hologram. In CDI, the resolution is given by the numerical aperture (NA) of the experimental setup:

$$R_{\text{CDI}} = \frac{\lambda}{2\sin\vartheta}, \tag{7}$$

where $\vartheta$ is the largest angle over which the scattered wave can be detected. Equation (7) represents the diffraction-limited resolution, as first introduced by Ernst Abbe [11, 12]. At small values of $\vartheta$, the approximations $\sin\vartheta \approx \frac{S}{2Z}$ and $R_{\text{CDI}} = \frac{\lambda}{2\sin\vartheta} \approx \frac{\lambda Z}{S} = R_{\text{Holography}}$ are valid. Here, the effective size of the hologram (diffraction pattern) is considered, which corresponds to the area over which interference can be observed.

## 1.3 Imaging of individual biological macromolecules: Radiation damage

Both X-ray and electron waves have sufficiently short wavelengths to give atomic resolution, and most of the imaging schemes in Fig. 1 have been successfully applied to the imaging of material science samples. It must be pointed out, however, that when imaging biological samples, the factor limiting the resolution is not the wavelength but the radiation damage [13]. Depending on the resolution required, the threshold dose of radiation for macromolecule imaging with high-energy electrons varies between 5 and 25 e/Å$^2$ to achieve a resolution of below 10 Å [14]. For X-rays, the radiation damage problem is even more severe, as inelastic scattering predominates over elastic scattering events, while only elastic scattering carries structural information [15]. For example, at a wavelength of 1 Å, the photoelectric cross-section of carbon is about 10 times higher than its elastic-scattering cross-section, making the photoelectric effect the primary source of damage. Depending on the desired resolution, the required dose of radiation damage can be approximated by the dependency dose (Gy) = $10^8 \times$ resolution (nm) [13]. Worldwide developments in XFELs have raised hopes of circumventing this undesirable ratio between elastic and inelastic scattering by employing extremely short, bright pulses, which allow the molecule to be imaged before it deteriorates [16]. It has been demonstrated that low-energy electrons (of kinetic energy 60–250 eV) can be employed to directly visualise individual biomolecules [17]; individual DNA molecules can withstand low-energy electron radiation of 60 eV energy (corresponding to a wavelength of about 1.6 Å) for 70 min, or a total radiation dose of $10^6$ e/Å$^2$, when imaged at a resolution of about 1 nm [17].

## 2. Survey of interferometric/coherent imaging schemes

### 2.1 Gabor in-line holography and point projection microscopy

Holography was invented by Dennis Gabor in 1947 [18-20]. The original experimental scheme proposed by Gabor was the same as the scheme shown in Fig. 1(a), where the reference wave and the scattered wave share the same optical axis. This type of holography is therefore called *in-line* Gabor holography [21, 22], as opposed to *off-axis* holography in which the reference and the object waves impinge on the recording plane at different angles. A similar experimental arrangement is sometimes called point projection microscopy (PPM) [23-31], an imaging scheme proposed by Morton and Ramberg in 1939 [32]. For an identical experimental arrangement, either a projection image or hologram of the sample can be observed at the detector depending on the parameters, and this difference defines the data analysis used and the name of the technique.

In holography, the unknown wave scattered by an object $O(\vec{\rho})$ is superimposed with a known reference wave $R(\vec{\rho})$. The resulting pattern, a hologram $H(\vec{\rho})$, captures the phase distribution of the scattered object wave [19, 20]:

$$H(\vec{\rho}) = \left|R(\vec{\rho}) + O(\vec{\rho})\right|^2 = \left|R(\vec{\rho})\right|^2 + O(\vec{\rho})R^*(\vec{\rho}) + O^*(\vec{\rho})R(\vec{\rho}) + \left|O(\vec{\rho})\right|^2, \tag{8}$$

where $\vec{\rho} = (X, Y, Z)$ is the coordinate in the hologram plane. Thus, the holography technique unambiguously solves the phase problem in a single step, due to the presence of the reference wave.

Reconstruction of a Gabor in-line hologram is performed numerically by multiplying the hologram with the reference wave $R(\vec{\rho})$, resulting in $R(\vec{\rho})H(\vec{\rho}) \propto \left|R(\vec{\rho})\right|^2 O(\vec{\rho}) \propto O(\vec{\rho})$, and by calculating the propagation of the complex-valued optical wave from the hologram plane backwards to the position of the object (based on Huygens principle and the Fresnel formalism):

$$u(\vec{r}) = \frac{i}{\lambda} \iint H(\vec{\rho}) R(\vec{\rho}) \frac{\exp(-ik|\vec{r} - \vec{\rho}|)}{|\vec{r} - \vec{\rho}|} d\sigma_S, \tag{9}$$

where $\vec{r} = (x, y, z)$ is the coordinate in the object plane, as illustrated in Fig. 2, and the integration is performed over the hologram plane. The result of this integral transform is a complex-valued distribution of the scattered object wave at any coordinate $\vec{r}$, and thus a *three-dimensional* (3D) distribution of the object scattered wave.

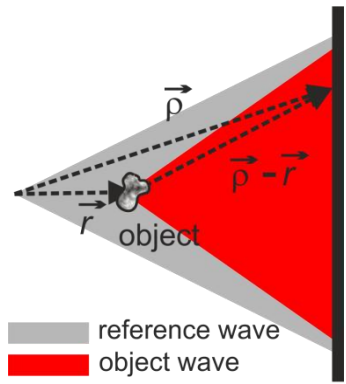

Fig. 2. Schematic of the back-propagation integral and three-dimensional reconstruction of an object from its Gabor in-line hologram.

In theory, the resolution achievable in digital Gabor in-line holography is given by Eq. (6); in practice, however, the resolution of in-line holography is limited by the visibility of the finest interference fringes, which are formed by interference between the reference wave and the object wave when scattered at large diffraction angles. Since any lateral shift of the sample is linearly translated into a shift of the hologram, even the smallest lateral shifts (which are unavoidable in experiments) blur these fine interference fringes. As a result of this smearing of the fringes, the resolution is worsened. Thus, in practice, the resolution of in-line Gabor holography is defined by the mechanical stability of the optical system [33]. One of the methods for increasing the resolution consists of laterally shifting the detection system during acquisition of the hologram, followed by alignment of the sequence of the recorded holograms using subpixel registration methods [34]. This allows for an increase in the number of pixels $N$ and a decrease in the pixel size $\Delta_H$, thus increasing the resolution, as for example demonstrated in [35].

Gabor in-line holography has been successfully applied in the imaging of biological macromolecules with low-energy electrons (50–250 eV). In-line holograms of individual molecules such as purple protein membrane [36], DNA molecules [17, 22, 25, 37, 38], phthalocyaninato polysiloxane molecule [23], the tobacco mosaic virus [39, 40], a bacteriophage [41], ferritin [42] and individual proteins (bovine serum albumin, cytochrome C and haemoglobin) [43] have been recorded and reconstructed. Despite the very short wavelength (0.8–1.7 Å) of the probing wave, the resolution of the imaged objects remains on the order of a nanometre, due to the mechanical stability of low-energy electron microscopes.

## 2.2 Coherent diffractive imaging (CDI)

CDI, which is sketched in Fig. 1(b), is similar to crystallographic experiments in which the sample is placed into a parallel beam and the scattered wave is recorded in the far field, but instead of a crystal, a non-crystalline object such as a single macromolecule is imaged. The scattered wave in the far field is described by the Fraunhofer diffraction regime, and is given by a Fourier transform (FT) of the exit wave $o(\vec{r})$, i.e. the wavefront distribution immediately behind the sample. The recorded intensity or diffraction pattern is given by:

$$I(\vec{k}) = \left| \iint o(\vec{r}) \, \exp(-i\vec{k}\vec{r}) \mathrm{d}\vec{r} \right|^2, \tag{10}$$

where $\vec{k} = (k_x, k_y, k_z)$ is the coordinate in the far field. The advantage of CDI is that the diffraction pattern is insensitive to lateral shifts of the sample, since these shifts only affect the phase distribution and not the recorded intensity. This allows for the preservation of the high-order diffraction signal. The disadvantage of CDI is that the phase of the scattered wave is completely lost and must be recovered.

### 2.2.1 Oversampling condition

In 1952, Sayre proposed a technique for recovering a crystal structure from its X-ray diffraction pattern alone, provided that the latter is sampled at such a fine rate (oversampled) that the intensity distribution between the Bragg peaks is available [44]. In 1972, Gerchberg and Saxton, who worked with transmission electron microscope (TEM) images, proposed an iterative algorithm to recover the complex-valued scattered object wave from its two amplitude measurements, i.e. at the object plane and the far-field plane [45]. In 2003, Miao et al. [46] combined these two ideas to successfully recover a non-crystalline object from its oversampled X-ray diffraction pattern. They demonstrated that the iterative algorithm converges (after several thousand iterations) if the initial conditions are such that the surroundings of the object ("support") are known. The experimental recording of an oversampled diffraction pattern followed by numerical recovery of the missing phase (and therefore the object structure) constitutes a novel class of high-resolution techniques called CDI.

The oversampling condition is illustrated in Fig. 3 and can be explained as follows. A two-dimensional (2D) continuous complex-valued distribution $o(x, y)$ is sampled in real space with $N \times N$ pixels, giving $o_1(m, n)$, where $m$ and $n$ are the pixel indices. The FFT of $o_1(m, n)$ is a complex-valued 2D distribution $O_1(p, q)$, also sampled with $N \times N$ pixels, where $p$ and $q$ are the pixel indices, as illustrated in the top row of Fig. 3.

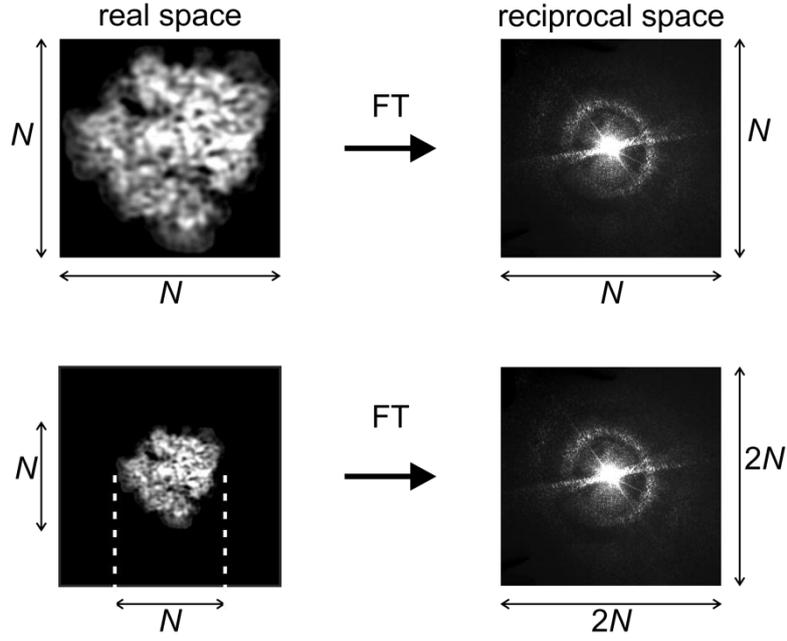

Fig. 3. Oversampling condition for an oversampling ratio $\sigma$ = 2. Sampling is shown at the Nyquist rate (upper row) and twice the Nyquist rate (lower row). Top row: The Fourier transform (FT) of a complex-valued object distribution filling the entire area and sampled with N × N pixels gives the spectrum sampled at exactly the Nyquist rate. Bottom row: The FT of a zero-padded complex-valued object distribution sampled with 2N × 2N pixels gives the spectrum sampled at twice the Nyquist rate ($\sigma$ = 2).

The sampling of the diffraction pattern in this case is done at the Nyquist rate, which can be described as follows. A 2D continuous complex-valued distribution $o(x, y)$ and its 2D complex-valued FT $O(u, v)$ are related by the analytical FT as:

$$O(u,v) = \iint o(x, y) \, \exp\left[-i2\pi(xu+yv)\right] \mathrm{d}x\mathrm{d}y. \tag{11}$$

When digitising 2D signals, continuous signals are replaced with their values at discretely selected coordinates, given by:

$$\begin{aligned} x &\to \Delta_x m, \quad m = 1...N \\ y &\to \Delta_y n, \quad n = 1...N \\ u &\to \Delta_u p, \quad p = 1...N \\ v &\to \Delta_v q, \quad q = 1...N. \end{aligned} \tag{12}$$

For the digitised signals, Eq. (11) becomes

$$O_1(p,q) = \sum_{m,n}^{N} o_1(m,n) \exp\left[-2\pi i \Delta_x \Delta_u (mp+nq)\right] = \sum_{m,n}^{N} o_1(m,n) \exp\left[-\frac{2\pi i}{N}(mp+nq)\right] \tag{13}$$

where we assume $\Delta_x = \Delta_y$, $\Delta_u = \Delta_v$, and the last part of Eq. (13) is the definition of the FFT. Thus, the FT of the digitally sampled signal can be calculated as the FFT when the following condition is fulfilled:

$$\Delta_x \Delta_u = \frac{1}{N}. \qquad (14)$$

This condition can be re-written as $\Delta_u = \frac{1}{N\Delta_x} = \frac{1}{S}$, where $S \times S$ is the sample area size. The sampling theorem states [47, 48]: "If a function $x(t)$ contains no frequencies higher than $B$ hertz, it is completely determined by giving its ordinates at a series of points spaced $1/(2B)$ seconds apart." In this theorem, $2B$ is the entire range of the signal spectrum. When the sampling theorem is applied to the sample and its diffraction pattern, the domains are swapped: the sample distribution is a finite distribution with extent $S$ (finite spectrum), and its FT is the distribution to be sampled (signal). Thus, the sampling of the diffraction pattern is done at the Nyquist rate $\Delta_u = \frac{1}{S} = \frac{1}{2B}$.

Next, 2D distribution $o_1(m, n)$ is zero-padded in real space up to $\sigma N \times \sigma N$ pixels, giving $o_2(m, n)$, where $\sigma$ is the oversampling ratio, $\sigma > 1$. The pixel size in the sample plane remains unchanged, $\Delta_x$. The FFT of $o_2(m, n)$ gives $O_2(p, q)$, which is sampled with $\sigma N \times \sigma N$ pixels, where the pixel size according to Eq. (14) is $\Delta_u = \frac{1}{\sigma N \Delta_x} = \frac{1}{\sigma S}$. Thus, $O_2(p, q)$ is sampled with smaller pixels than $O_1(p, q)$ but the extent of $O_2(p, q)$ and $O_1(p, q)$ is the same and given by $\sigma N \Delta_u = \frac{1}{\Delta_x}$. This means that the distribution of the FT spectrum of the zero-padded signal is the same as that of the non-zero-padded signal, but is sampled with more pixels ($\sigma N \times \sigma N$) of smaller size ($\Delta_u = 1/\sigma S$), as illustrated in the bottom row of Fig. 3. Thus, when diffraction pattern of an object is sampled at a rate higher than the Nyquist rate, it is equivalent to zero-padding of the object, and vice-versa. The presence of the zeros in the object domain, in turn, helps in solving the following set of equations for unknown phases:

$$I(p,q) = |O_2(p,q)|^2 = \left| \sum_{m,n=1...N} o_2(m,n) \exp\left(-\frac{2\pi i}{N}(mp+nq)\right) \right|^2, \qquad (15)$$

which originate from the fact that the diffraction pattern is the squared amplitude of the object exit function. The real and imaginary parts of the object distribution $o_1(m, n)$, each sampled with $N \times N$ pixels, give rise to $2N^2$ unknowns. When Eq. (15) is written for each pixel ($p$, $q$) in the detector domain, this gives rise to ($\sigma N$)$^2$ equations. Thus, in order for the system of equations Eq. (15) to have a unique solution, the number of unknowns must be equal to the number of equations, which is achieved when the oversampling ratio is $\sigma = \sqrt{2}$ in each sampling direction. In practice, the oversampling ratio $\sigma$ is often selected such that $\sigma > 2$, to ensure faster convergence of the iterative phase retrieval routine [49].

### 2.2.2 Iterative phase retrieval

Unlike in holography, no reference wave is involved in CDI, and the reconstruction is not straightforward. Instead, the object distribution is retrieved from its diffraction pattern using an iterative phase retrieval procedure. Most of the effective phase retrieval methods are based on error reduction (ER) and hybrid input–output (HIO) algorithms, as described by Fienup in 1982 [50].

A typical iterative reconstruction loop is sketched in Fig. 4. It includes the following steps:

(i) Formation of a complex-valued field $O(u,v)$ at the detector plane. The amplitude is always given by the square root of the measured intensity $O_0(u,v)$.

(ii) Propagation of the wavefront from the detector plane to the sample plane by calculating the inverse FT of $O(u,v)$. The result is the object distribution $o(x,y)$.

(iii) Application of constraints to the object distribution $o(x,y)$. The object must be zero-padded, that is, it must occupy a certain limited area, which is regulated by masking that area. Other constraints are applied to the object transmission function; for instance, for X-ray diffraction images, the reconstructed electron density must be real and positive.

(iv) Calculation of the FT of the updated object distribution $o'(x,y)$. This gives rise to the wavefront distribution in the detector plane $O'(u,v)$. The phase of the $O'(u,v)$ distribution is adapted for use in step (i) in the next iteration.

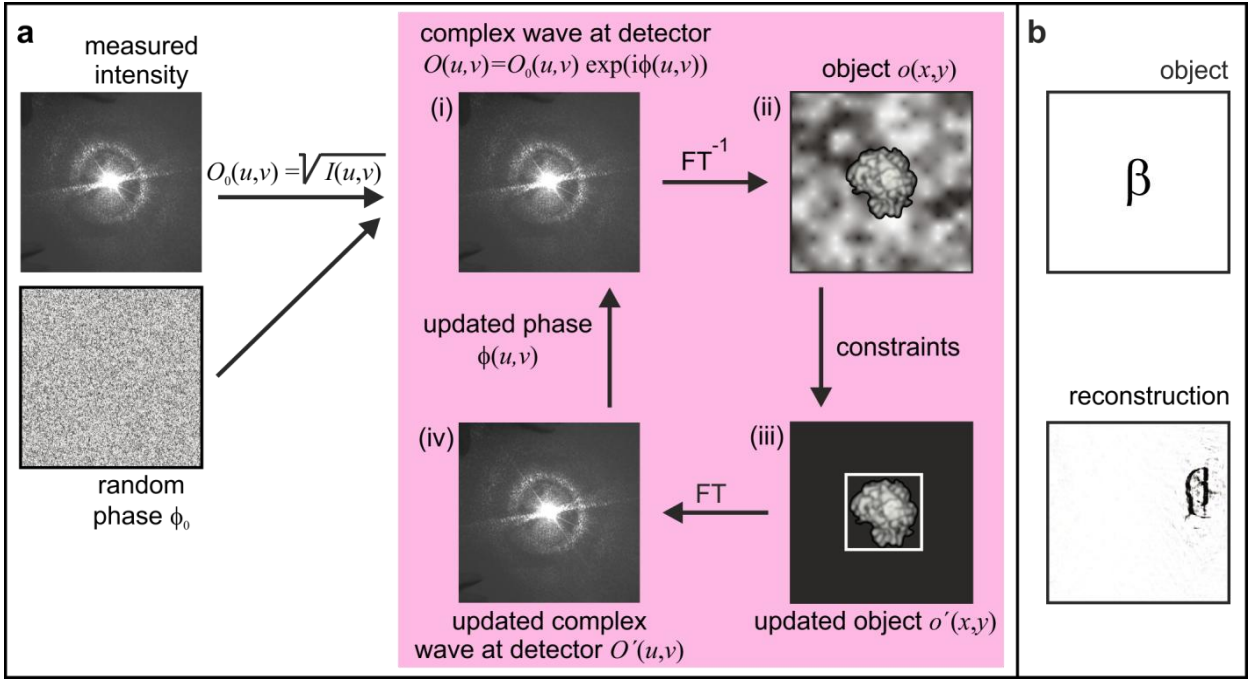

Fig. 4. Iterative reconstruction of coherent diffraction images. (a) The iterative loop includes steps (i)–(iv). In the first iteration, the phase distribution in the detector plane is randomly distributed. (b) Example of a misleading reconstruction of the diffraction pattern of β (obtained using a HIO algorithm with feedback parameter 0.9).

To monitor the convergence of the algorithm, the error can be calculated as the mismatch between the amplitude in the detector plane, updated after each iteration, and the measured amplitude [50]:

$$\text{Error} = \frac{\sum_{u,v} \left\| |O_0(u,v)| - |O'(u,v)| \right\|}{\sum_{u,v} |O_0(u,v)|}, \qquad (16)$$

where the summation is performed over the pixels in the detector plane. Alternatively, the error is calculated by estimating the object distribution and updating this after each iteration [51-53]:

$$\text{Error} = \frac{\sum_{(x,y) \in \text{Support}} |o(x,y)|^2}{\sum_{(x,y) \notin \text{Support}} |o(x,y)|^2}, \qquad (17)$$

where the summation is performed over the pixels in the object domain.

The intrinsic resolution in CDI is defined by Eq. (7). When arranging an experimental CDI setup, two parameters must be considered: (i) the oversampling ratio, which must exceed two, i.e. $\sigma > 2$, which sets limits on the pixel size in the detector domain and the number of pixels; (ii) the achievable resolution of the reconstructed object, which is given by Eq. (7). The resolution of the reconstructed object can be estimated by calculating the phase retrieval transfer function (PRTF) [54, 55]:

$$\text{PRTF}(\xi) = \frac{|O(u,v)|}{O_0(u,v)}, \tag{18}$$

which is usually represented in form of a one-dimensional angular-averaged PRTF as a function of the frequency $\xi = \sqrt{u^2 + v^2}$ in the Fourier domain. The purpose of the PRTF is to evaluate the recovered phases in the detector domain by comparing the recovered amplitudes with the measured amplitudes. The phases which are recovered with less consistency have a smaller PRTF, and the resolution cutoff is given by the spatial frequency at which the PRTF extrapolates to zero.

There are a number of problems associated with numerical reconstruction in CDI [49]. (i) The solution may be ambiguous, an example of a typically misleading result of phase retrieval using the HIO algorithm is shown in Fig. 4(b). In general, the results of hundreds of iterative runs are averaged to arrive at a correct reconstruction. (ii) The iterative process may stagnate at partial solutions. (iii) Phase retrieval is only possible if the diffraction pattern is oversampled. Thus, the geometry of the experimental setup must be designed to fulfil the oversampling condition, rather than allowing measurement of the highest possible diffraction angle and thus the highest resolution. (iv) The oversampling condition in the detector plane corresponds to zero-padding in the object plane, which requires the sample to be surrounded by a support with known transmission properties. (v) Signals in the central overexposed region of the diffraction pattern may be missing. The intensity ratio between the central spot and the signal at the rim of the detector can reach values of $10^7$; commonly used 16-bit cameras are simply not capable of capturing the entire intensity range, and the central (low-resolution) part is usually sacrificed by being blocked. These missing data are usually obtained by recording a low-resolution image using some other technique, for example transmission electron microscopy [56], or by recording a set of images at different exposure times [54, 57]. Due to these shortcomings, there are ongoing searches for alternative experimental techniques and better reconstruction methods [58, 59].

Despite these drawbacks, the power of the CDI technique has been demonstrated by reconstructing the structure of a double-walled carbon nanotube at a resolution of 1 Å, from a coherent diffraction pattern recorded using a transmission electron microscope operating at 200 keV and exhibiting a nominal point resolution of 2.2 Å [56]. Thus, the achieved resolution is *the highest possible resolution* approaching the resolution obtained in crystallographic experiments.

Biological specimens have been imaged by CDI using coherent X-rays [54, 60-67]. Overviews of CDI applications are presented in references [68, 69], and some results are shown in Fig. 5. Despite the very short wavelength applied, the resolution remains on the order of nanometres due to the radiation damage problem. The X-ray diffraction pattern of a crystal, unlike that of an individual molecule, displays a strong signal due to the periodicity of the crystal, and obtaining the X-ray diffraction pattern of *an individual molecule* requires a much more intense X-ray beam. As a consequence, the resolution is limited by radiation damage, and remains very moderate. A few biological specimens have been imaged by CDI using X-rays at a resolution of a few nanometres, for example an unstained yeast cell with resolution 30 nm [54], E. coli bacteria with resolution 30 nm [60], single herpes virions with resolution 22 nm [61], malaria-infected red blood cells with resolution 40 nm [62], a frozen hydrated yeast cell with resolution 25 nm [63], human chromosomes with resolution 38 nm [64], labelled yeast cells [65], and deinococcus radiodurans bacterial cells with resolution 85 nm [66] and 50 nm when imaged three-dimensionally [67].

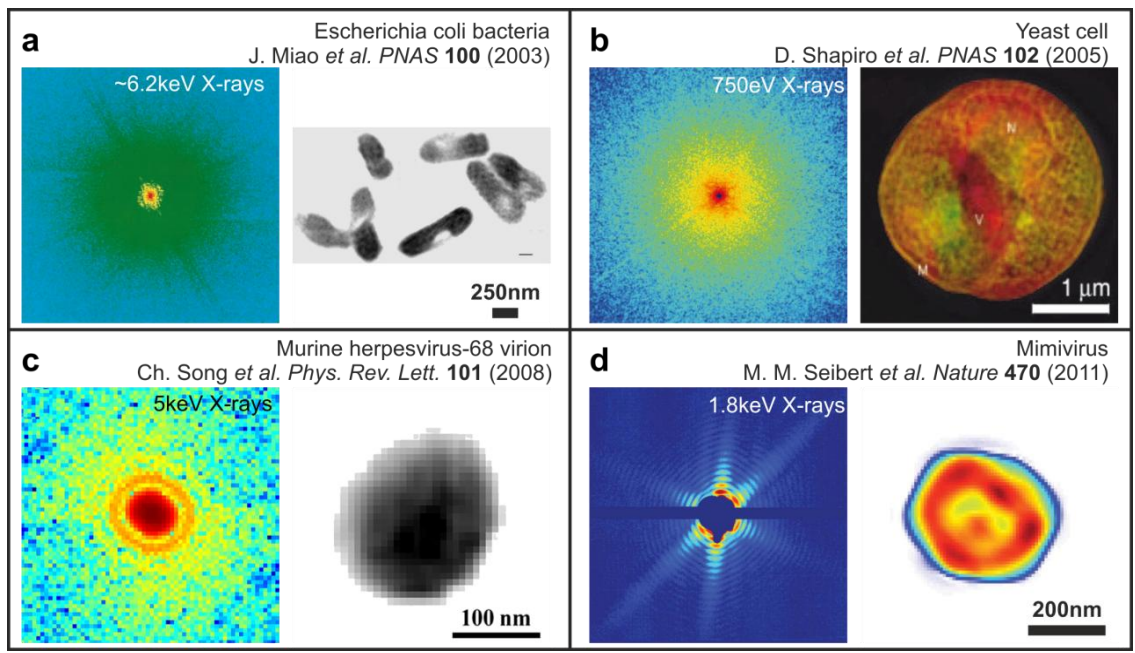

Fig. 5. Overview of the imaging of biological molecules using X-ray CDI. (a) Escherichia coli bacteria [60], copyright (2003) National Academy of Sciences. (b) Unstained yeast cell [54], copyright (2005) National Academy of Sciences. (c) Murine herpesvirus-68 virion, reprinted from [61], copyright (2008) by the American Physical Society. (d) Mimivirus, reprinted from [70] by permission from Springer Nature, copyright 2011.

Ultra-short and extremely bright coherent X-ray pulses from XFEL allow for recording of a high-resolution diffraction pattern before the sample explodes [16, 71]. The initial results from the first XFEL facility to be operational (the Linac Coherent Light Source) reported the imaging of an individual unstained mimivirus with resolution 32 nm [70]; in this experiment, an X-ray pulse with 1.8 keV (6.9 Å) energy and 70 fs duration was focused on a spot 10 μm in diameter with $1.6 \times 10^{10}$ photons per 1 μm$^2$. A sub-nanometre resolution could be achieved by employing shorter pulses and a higher photon flux [70, 71]; although at present this is beyond the abilities of XFEL, it may be realised with the next generation of systems. At present, single-particle imaging (SPI) using XFELs is realised as follows: (a) hundreds or thousands of diffraction patterns of identical molecules are acquired; (b) these diffraction patterns are assigned to different orientations of the molecule (classes) using methods adapted from cryo-electron microscopy; (c) the 3D diffraction pattern is built, which is (d) reconstructed using iterative phase retrieval methods. SPI has been successfully applied to the imaging of individual mimivirus [72] and microtubule molecules (with resolution 2 nm) [73].

## 2.3 Fourier Transform Holography (FTH)

The scheme used for Fourier transform holography (FTH) [74] is sketched in Fig. 1(c) and shown in more detail in Fig. 6. The arrangement is the same as in CDI, but with one important difference: a small aperture or a scatterer is placed next to the object to provide a source for the reference wave. The experimental FTH record is therefore in fact a diffraction pattern with a superimposed spherical reference wave. The presence of the reference wave ensures that the phase distribution of the scattered object wave is captured, similarly to holography, which simplifies the reconstruction procedure. The reference wave in FTH is typically created by diffraction from a point-like aperture [74-76].

The sample distribution is reconstructed from its FTH hologram by simply calculating the FT of the hologram. According to the Wiener-Chintchin theorem [77, 78], the FT of the power spectrum of a function is the auto-correlation of the function:

$$I(u,v) = \left| \mathrm{FT}[t(x,y)] \right|^2, \quad \mathrm{FT}[I(u,v)] = \iint t(\mu,\eta)\, t^*(\mu-x, \eta-y)\mathrm{d}\mu \mathrm{d}\eta = t(x,y) \circ t(x,y), \quad (19)$$

where $(u, v)$ are the coordinates in the detector plane and ∘ denotes correlation. If the sample distribution described by the transmission function $t(x, y)$ includes a tiny hole (or alternatively a point scatterer, although this is more difficult experimentally) in addition to the main object described by the $o(x, y)$ function, the total transmission function of the sample can be written as:

$$t(x, y) = o(x, y) + \delta\left(x - x_0, y - y_0\right), \quad (20)$$

where $(x_0, y_0)$ is the position of the pinhole. The autocorrelation of such a function includes the following terms:

$$t(x,y) \circ t(x,y) = \delta(x,y) + o(x,y) \circ o(x,y) + o(x+x_0, y+y_0) + o^*(-x+x_0, -y+y_0). \qquad (21)$$

In Eq. (21), the first term is the sharp intensity peak at the centre, the second term is the auto-correlation of the object, also located in the centre and about twice the size of the object, and the final terms are two mirror-symmetric images of the object placed at the position of the pinhole (see Fig. 6(c)). An unambiguous reconstruction of the object can thus be obtained simply by taking the FT of the FTH record. FTH has been successfully realised with visible light [57] and X-rays [75, 76, 79].

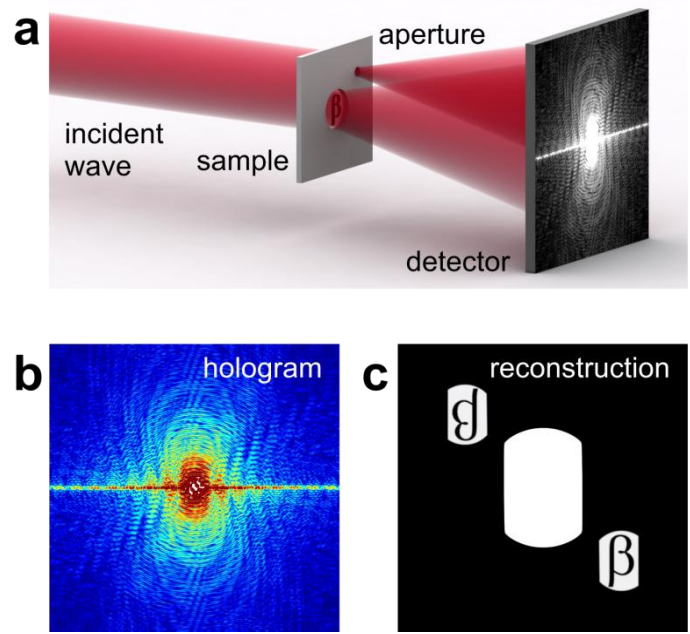

Fig. 6. Principle of operation of FTH. (a) Experimental scheme for FTH. (b) Fourier transform hologram. (c) Sample distribution reconstructed by taking the FT of the hologram and exhibiting two mirror-symmetric reconstructions at the aperture position.

The resolution of the reconstructed object in FTH is determined by the size of the reference source (aperture or scatterer), which in reality is not an ideal mathematical $\delta$-function but has a finite size. Although a small reference source is preferred, the creation of small apertures (sub-nanometre sized in the case of electron or X-ray waves) is a practical challenge. The smaller the reference source, the lower the intensity of the reference wave, and in order to obtain a good contrast diffraction pattern, the intensities of the reference and the object waves should be approximately equal. Several solutions to this dilemma have been proposed, which involve using various forms of the reference source: an extended reference source [80-82], multiple reference sources [83, 84], and a structured reference source [79, 85]. The initial reconstruction, obtained only from an FT of the hologram, can be further refined by an iterative procedure similar to that shown in Fig. 4, which allows for a resolution approaching that achievable in CDI [86].

## 2.4 Fresnel Coherent Diffractive Imaging (FCDI)

Fresnel coherent diffraction imaging (FCDI) [87-90], as shown in Fig. 1(d), is another attempt to introduce a holographic component to CDI. In FCDI, the incident wave is slightly divergent (1–2°), which results in the formation of an in-line hologram in the centre of the diffraction pattern, as illustrated in Fig. 7. The accessible holographic and thus phase information allows for an immediate low-resolution reconstruction of the object and an estimate of the phase of the object scattered wave. The available phase distribution is then plugged into the iterative phase retrieval loop shown in Fig. 4, ensuring fast convergence of the iterative process and the stability of the solution. There are several drawbacks to this method, however. Similarly to the diffraction patterns in CDI, the average intensity in the central "holographic" region is about $10^4$ times higher than the intensity of the "diffraction" part of the pattern. Hence, to record both holographic and CDI parts in the same image, a set of images is required that are obtained at different exposure times (including a very long exposure to record the signal at the rim of the detector). Since the wavefront in FCDI is not plane but spherical, the wavefront distribution in the far field is not simply the FT of the exit function, and is highly sensitive to any lateral shifts, in the same way as in Gabor in-line holography. Although a longer acquisition time is expected to increase the contrast of higher-order diffraction signals in the diffraction pattern, in fact it just blurs out the high-order diffraction information.

The sphericity of the incident wavefront causes higher-order diffraction to occur at larger angles than when the incident beam is planar, and some resolution is therefore lost. The resolution in FCDI, as in CDI, is defined by Eq. (7), where $\vartheta$ is the largest angle at which the scattered wave can be detected.

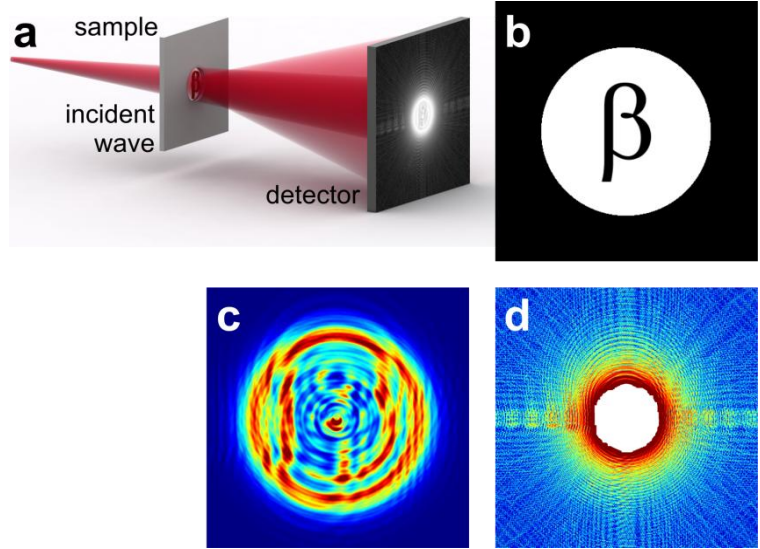

Fig. 7. Principle of operation of FCDI. (a) Experimental arrangement. (b) Sample. (c) Central part of the Fresnel coherent diffraction image, which contains holographic information. (d) Outer part of the Fresnel coherent diffraction image, which contains higher-order diffraction information. Note that the outer part becomes visible when the central holographic part is overexposed.

## 2.5 Ptychography

Ptychography [91-96], which is illustrated in Fig. 1(e), relies on lateral shifts of the sample, or more conventionally of the probing wave. When the probing beam is shifted, the exit wave is given by:

$$\psi_j(x,y) = o(x,y) P(x-X_j, y-Y_j), \qquad (22)$$

where $P(x-X_j, y-Y_j)$ is the probing wavefront and $(X_j, Y_j)$ is the relative shift between the object and the probing beam during recording of the $j$-th diffraction pattern. A sequence of diffraction patterns for different shifts of the probing beam is recorded. The reconstruction uses an iterative procedure, and the steps used in the extended ptychographical iterative engine (ePIE) algorithm [97, 98] are listed below. At the beginning of the reconstruction process, the object distribution $o(x,y)=1$ is assumed, and the probing wavefront function $P(x,y)$ is guessed. For each position $j = 1,...,J$ of the probing beam, the corresponding diffraction pattern $I_j(u,v)$ is assigned.

(i) For each shifted probing wave $P(x-X_j, y-Y_j)$, the object distribution $o(x,y)$ is multiplied with the incident wave, thus giving a current guess for the exit wave function

$$\psi_g(x,y) = P(x-X_j, y-Y_j) o(x,y).$$

(ii) The FT of the obtained exit wave $\psi_g(x,y)$ is calculated, giving a complex-valued wavefront distribution in the detector plane, $\Psi_g(u,v)$.

(iii) The amplitude distribution of the wavefront obtained at the detector, $\Psi_g(u,v)$, is replaced by the measured amplitudes, giving a corrected wavefront distribution in the far field, $\Psi_c(u,v)$.

(iv) The inverse FT of $\Psi_c(u,v)$ gives a corrected exit function $\psi_c(x,y)$.

(v) An updated estimate of the object function is then computed according to

$$o_{n+1}(x,y) = o_n(x,y) + \alpha \frac{\left[P(x-X_j, y-Y_j)\right]^*}{\left|P(x-X_j, y-Y_j)\right|^2_{\max}} \left[\psi_c(x,y) - \psi_g(x,y)\right],$$

where $\alpha$ is a constant (typically $\alpha$ = 1). An updated estimate of the probe function is computed according to

$$P_{n+1}(x,y) = P_n(x,y) + \beta \frac{\left[o_n(x+X_j, y+Y_j)\right]^*}{\left|o_n(x+X_j, y+Y_j)\right|^2_{\max}} \left[\psi_c(x,y) - \psi_g(x,y)\right],$$

where $\beta$ is a constant (typically $\beta = 1$).

The steps set out above are repeated for each $j$, and this completes one iteration of the ePIE.

The resolution of the reconstructed sample distribution can be enhanced by incorporating the extrapolation of the acquired diffraction patterns into the reconstruction algorithm [99]. 2D ptychographic measurements can be utilised to reconstruct a 3D sample distribution using multi-slicing calculations for the propagation of the wavefront through the 3D sample, and this approach is known as 3D ptychography [100, 101].

Ptychography has been realised with light [97, 99, 102], X-rays [66, 67, 103-110], electrons [98, 111], terahertz waves [112], and other types of radiation [113]. Ptychography can be implemented with any wavefront, for example with a divergent probing wave in the FCDI regime [104, 108, 109], and can also be combined with tomography [67, 105, 108, 109, 114]. In ptychography, the resolution is determined by the resolution of the employed imaging mode. For example, with a plane probing wave in CDI mode, the resolution is the same as in CDI, and is given in Eq. (7). X-ray ptychography has been applied to the imaging of biological objects [105, 106] and individual bacterial cells at a resolution of 85 nm [66].

# 3. Development of in-line holography and CDI with low-energy electrons

This section describes some of the experimental work advancing holography and CDI techniques, and in particular the imaging of individual biomolecules with low-energy electrons carried out by Professor Hans-Werner Fink's group at the University of Zurich. To test different imaging schemes, a dedicated optical setup was built and reconstruction algorithms were designed and tested on optical coherent images. These algorithms were then used to reconstruct images of samples acquired with low-energy electrons.

## 3.1 Experimental setups

A light optical setup which can be used for all the imaging schemes shown in Fig. 1 was built in order to test different imaging modes and phase retrieval techniques. Examples of two such optical schemes, holography and CDI, are shown in Fig. 8. The light optical coherent diffraction experiments discussed below were all performed using this setup. Since the mechanism of photon scattering is the same as for visible light or X-ray waves, imaging methods developed and demonstrated with light can be employed directly in X-ray diffraction experiments.

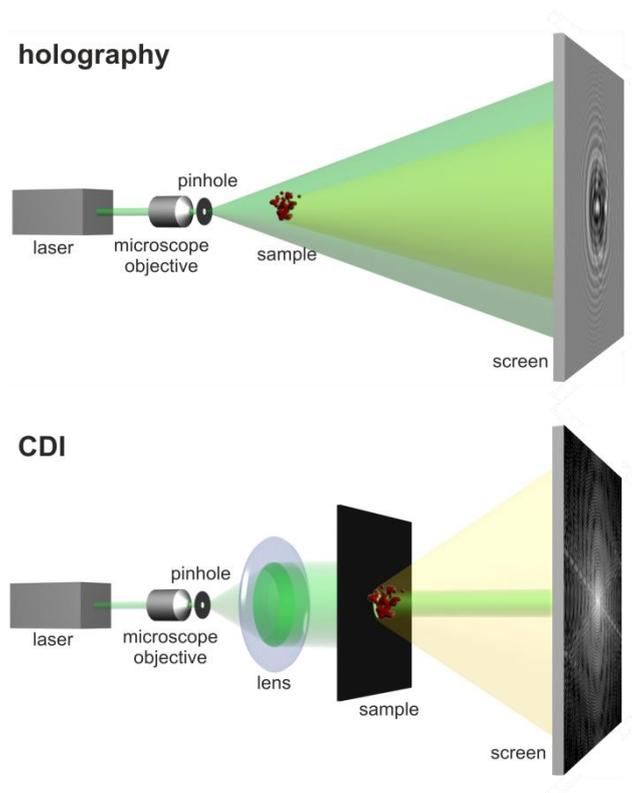

Fig. 8. Optical schemes for holography and coherent diffractive imaging (CDI). Laser light with wavelength 532 nm (diode pumped all-solid-state laser from Changchun New Industries Optoelectronics Tech. Co.) was spatially filtered by the microscope objective and pinhole assembly (Newport Three Axis Spatial Filter with M60x NA = 0.85 microscope objective). The divergent wavefront was collimated with a lens (Thorlabs) for the CDI experiments. The screen, which was made of semi-transparent Mylar-like material, could be positioned at any distance behind the sample. This allowed images of sizes ranging from 22 × 22 mm$^2$ to about 360 × 360 mm$^2$ to be captured. The camera (Hamamatsu C4742-95) was operated using self-designed LabView software.

Self-built electron microscopes for the experiments with coherent low-energy electrons were used by Professor Fink's group at the University of Zurich, and these are sketched in Fig. 9. In total, there were three low-energy electron microscopes with slightly different configurations, i.e. with source-to-detector distances 47, 68 and 180 mm and different detection systems. The source of the coherent electron beam was a sharp W(111) tip, and the electrons were extracted by field emission [115]. The position of the tip was controlled by a three-axis piezo-manipulator with nanometre precision. The modified low-energy electron microscope equipped with a microlens [116, 117], which could operate in both holography and CDI modes, is shown in Fig. 9 [118]. The experimental holograms and diffraction patterns, which were recorded with low-energy electrons and are shown below, were obtained using the microscopes sketched in Fig. 9. The low-energy electron microscopes employed in this study have been described in detail in previous publications [21, 22, 37, 41-43], and other low-energy electron microscopes used by other groups for in-line holography or PPM are described in references [29, 36, 119].

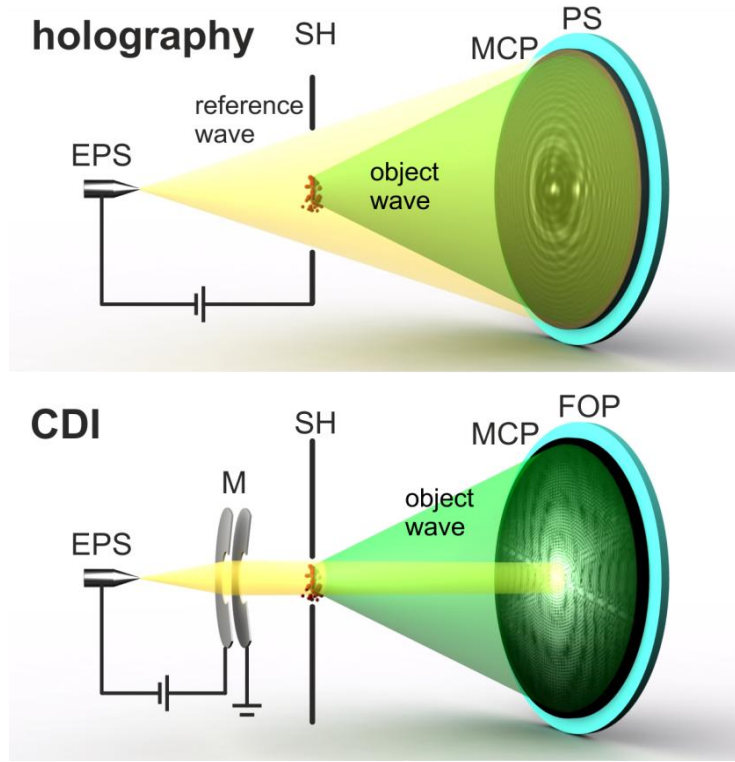

Fig. 9. Coherent low-energy electron microscopes. Coherent low-energy electrons are extracted from the electron point source (EPS) by field emission. The sample is fixed in the sample holder (SH). In the holographic microscope, the interference between the scattered and unscattered (reference) waves is recorded by the detection system, which consists of a micro-channel plate (MCP) with an adjacent phosphorous screen (PS). In the CDI microscope, the electron beam is collimated by a microlens (M) and the detection system consists of the MCP followed by a fibre optic plate (FOP) with a thin layer of phosphor.

## 3.2 Theory of formation and reconstruction of in-line holograms

### 3.2.1 Transmission function of sample

In general, the interaction of a wave with a medium is described by a complex-valued transmission function [120]:

$$t(x, y) = \exp[-a(x, y)]\exp[i\varphi(x, y)], \qquad (23)$$

where $a(x, y)$ is the object absorption distribution and $\varphi(x, y)$ is the object phase distribution, or the phase delay introduced by the object into the incident beam. The distribution of the wavefront immediately behind the sample is called the exit wave. For a plane incident wave, the exit wave is given by $t(x, y)$. The transmission function depends on the properties of both the probing radiation and the matter, and is generally complex-valued. In most cases, both the absorption and phase distributions are significantly different from zero, and approximations involving weak absorption or a weak phase object cannot be applied. Reconstruction of the absorption and phase properties constitutes a challenge in terms of the numerical retrieval of the object distribution, in all coherent imaging techniques.

### 3.2.2 Simulation and reconstruction of in-line holograms

The propagation of the wavefront $u_1(\vec{r}_1)$ from the plane described by coordinate $\vec{r}_1$ to the plane described by coordinate $\vec{r}_2$ is given by the integral transformation based on the Huygens-Fresnel principle:

$$u_2(\vec{r}_2) = -\frac{i}{\lambda} \iint u_1(\vec{r}_1) \frac{\exp(ik|\vec{r}_1 - \vec{r}_2|)}{|\vec{r}_1 - \vec{r}_2|} d\sigma, \tag{24}$$

where $\dfrac{\exp(ik|\vec{r}_1 - \vec{r}_2|)}{|\vec{r}_1 - \vec{r}_2|}$ are the secondary waves and the integration is performed over the plane defined by coordinate $\vec{r}_1$.

**Simulation.** The original in-line Gabor holography scheme employs a divergent spherical probing wavefront. We therefore mainly consider this type of wavefront. For spherical wavefront, the diffracted wave in the detector plane (far field) is described by the Fresnel diffraction integrals. The interference pattern (hologram) in this case has the same distribution at any distant detecting plane, and moving the detector along the optical axis changes only the magnification rather than the overall distribution of the hologram.

The complex-valued distribution of the diffracted wave is given by Eq. (24), where the distribution of the wavefront in the sample plane is given by

$$u_1(\vec{r}) = \frac{\exp(ikr)}{r} t(\vec{r}) , \tag{25}$$

where $\dfrac{\exp(ikr)}{r}$ is the probing divergent spherical wave and $t(\vec{r})$ is the transmission function. By substituting Eq. (25) into Eq. (24) we obtain:

$$U(\vec{\rho}) = -\frac{i}{\lambda} \iint \frac{\exp(ikr)}{r} t(\vec{r}) \frac{\exp(ik|\vec{r} - \vec{\rho}|)}{|\vec{r} - \vec{\rho}|} d\sigma, \tag{26}$$

where $\vec{\rho}=(X,Y,Z)$ and $\vec{r}=(x,y,z)$ are coordinates in the hologram and the object planes, respectively, and the integration in performed over the plane of the object's location. In the paraxial approximation, Eq. (26) describes the diffracted wave in the form of a convolution:

$$U(X,Y) \approx -\frac{i}{\lambda z(Z-z)} \iint_S \exp\left[\frac{i\pi}{\lambda z}(x^2+y^2)\right] t(x,y) \exp\left\{\frac{i\pi}{\lambda(Z-z)}\left[(x-X)^2+(y-Y)^2\right]\right\} dxdy \approx$$
$$\approx -\frac{i}{\lambda z(Z-z)} \exp\left[\frac{i\pi}{\lambda Z}(X^2+Y^2)\right] \iint_S t(x,y) \exp\left\{\frac{i\pi}{\lambda z}\left[\left(x-\frac{z}{Z}X\right)^2+\left(y-\frac{z}{Z}Y\right)^2\right]\right\} dxdy, \quad (27)$$

where the convolution is performed between the transmission function of the sample $t(x,y)$ and a Fresnel function $\exp\left[\frac{i\pi}{\lambda z}(x^2+y^2)\right]$, and the result of the convolution is expressed in the scaled coordinates $X'=\frac{z}{Z}X=\frac{X}{M}$ and $Y'=\frac{z}{Z}Y=\frac{Y}{M}$, where $M=\frac{Z}{z}$ is the magnification factor.

The analytical FT of the Fresnel function is given by:

$$\text{FT}\left\{\exp\left[\frac{i\pi}{\lambda z}(x^2+y^2)\right]\right\} =$$
$$\iint \exp\left[\frac{i\pi}{\lambda z}(x^2+y^2)\right] \exp\left[-2\pi i(xu+yv)\right] dxdy = i\lambda z \exp\left[-i\pi\lambda z(u^2+v^2)\right]. \quad (28)$$

Applying the FFT routine leads to the following sampling relationship and pixel size in the Fourier domain (see Eqs. (11)–(14)):

$$\Delta_x \Delta_u = \frac{1}{N}, \rightarrow \Delta_u = \frac{1}{N\Delta_x} = \frac{1}{S} \quad (29)$$

where $S\times S$ is the sample area size. By substituting Eq. (29) into Eq. (28), we obtain

$$\text{FT}\left\{\exp\left[\frac{i\pi}{\lambda z}(x^2+y^2)\right]\right\} = i\lambda z \exp\left[-\frac{i\pi\lambda z}{S^2}(p^2+q^2)\right], \quad (30)$$

where $p$ and $q$ are the pixel indices $p=1...N$, $q=1...N$. The FT of the Fresnel function should be calculated directly using Eq. (30), since with typical parameters, the function given by Eq. (30) can be correctly sampled, unlike the Fresnel function in real space. The factor $i\lambda z$ compensates for the factor $-i/(\lambda z)$ in Eq. (27).

The diffracted wave in the detector plane can be simulated using the following steps:
(i) Applying the FFT to $t(x,y)$;

(ii) Multiplying the result of (i) with the numerically calculated $\exp\left[-\frac{i\pi\lambda z}{S^2}(p^2+q^2)\right]$;

(iii) Applying the inverse FFT to the result of (ii).

The hologram is then calculated as

$$H(X,Y) = |U(X,Y)|^2. \tag{31}$$

The hologram distribution formed in this way can be interpreted as follows. The sample transmission function can be represented as:

$$t(x,y) = 1 + o(x,y), \tag{32}$$

where $o(x,y)$ is a perturbation caused by the presence of the object. The wavefront in the detector plane is given by the sum of the reference and object waves. The reference wave originates from the first term (i.e. 1) in Eq. (32); this corresponds to the probing wave, which when propagated to the detector plane is given by:

$$R(X,Y) = \frac{\exp(ik\rho)}{\rho}. \tag{33}$$

This can be quantitatively checked by substituting $t(x,y) = 1$ into the propagation integrals in Eq. (27). The object wave originates from the term $o(x,y)$ in Eq. (32), and is described by the distribution $O(X,Y)$ in the detector plane. The total wavefront in the detector plane is given by:

$$t(x,y) \to U(X,Y) = R(X,Y) + O(X,Y), \tag{34}$$

where the wave propagation indicated by the arrow is described by the integral in Eq. (27).

**Reconstruction.** Reconstruction is performed by multiplying the hologram with the reference wave

$$R(X,Y)H(X,Y) \propto |R(X,Y)|^2 O(X,Y) \propto O(X,Y), \tag{35}$$

followed by propagation of the resulting wavefront back to the object plane, which is calculated using Eq. (24):

$$u(\vec{r}) = \frac{i}{\lambda} \iint \frac{\exp(ik\rho)}{\rho} H(\vec{\rho}) \frac{\exp(-ik|\vec{r}-\vec{\rho}|)}{|\vec{r}-\vec{\rho}|} d\sigma_H, \tag{36}$$

where $\frac{\exp(ik\rho)}{\rho}$ is the reference wave (Eq. (33)), $H(\vec{\rho})$ is the hologram distribution, and the integration is performed over the hologram plane. In the paraxial approximation, Eq. (36) becomes:

$$u(x,y) \approx \frac{i}{\lambda Z(Z-z)} \iint_H \exp\left[\frac{i\pi}{\lambda Z}(X^2+Y^2)\right] H(X,Y) \exp\left\{-\frac{i\pi}{\lambda(Z-z)}\left[(x-X)^2+(y-Y)^2\right]\right\} dXdY \approx$$

$$\approx \frac{i}{\lambda Z^2} \exp\left[\frac{i\pi}{\lambda z}(x^2+y^2)\right] \iint_H H(X,Y) \exp\left\{-\frac{i\pi z}{\lambda Z^2}\left[\left(X-\frac{Z}{z}x\right)^2+\left(Y-\frac{Z}{z}y\right)^2\right]\right\} dXdY, \tag{37}$$

which is a convolution with the Fresnel function.

The analytical FT of the Fresnel function is given by:

$$\text{FT}\left\{\exp\left[-\frac{i\pi z}{\lambda Z^2}(X^2+Y^2)\right]\right\} = \iint \exp\left[-\frac{i\pi z}{\lambda Z^2}(X^2+Y^2)\right]\exp\left[-2\pi i(Xu+Yv)\right]dXdY = -\frac{i\lambda Z^2}{z}\exp\left[\frac{i\pi\lambda Z^2}{z}(u^2+v^2)\right]. \tag{38}$$

Applying the FFT routine leads to the following sampling relationship and pixel size in the Fourier domain (see Eqs. (11)–(14)):

$$\Delta_H \Delta_u = \frac{1}{N}, \rightarrow \Delta_u = \frac{1}{N\Delta_H} = \frac{1}{S_H}, \tag{39}$$

where $S_H \times S_H$ is the size of the hologram area, which is related to the area of the magnified object by

$$S_H = S\frac{Z}{z}. \tag{40}$$

By substituting Eq. (39) and (40) into Eq. (38), we obtain

$$\text{FT}\left\{\exp\left[-\frac{i\pi z}{\lambda Z^2}(X^2+Y^2)\right]\right\} = -\frac{i\lambda Z^2}{z}\exp\left[\frac{i\pi\lambda z}{S^2}(p^2+q^2)\right], \tag{41}$$

where $p$ and $q$ are the pixel indices $p=1...N$ and $q=1...N$. Equation (41) is the same convolution kernel as in the hologram simulation (Eq. (30)), although this time it is complex-conjugated. Also here, the FT of the Fresnel function should be calculated directly using Eq. (41), as in the simulation, since with typical parameters the function given in Eq. (41) can be correctly sampled, unlike the Fresnel function in real space. The factor $\left(-i\lambda Z^2/z\right)$ in Eq. (41) compensates for the factor $\left(i/\lambda Z^2\right)$ in Eq. (37).

Thus, the reconstruction of in-line hologram can be achieved using the following steps:
(i) Applying the FFT of $H(X,Y)$;

(ii) Multiplying the result of (i) with the numerically calculated $\exp\left[\frac{i\pi\lambda z}{S^2}(p^2+q^2)\right]$;

(iii) Applying the inverse FFT to the result of (ii).

The result of this reconstruction will be the distribution of the exit wave. In the case where no object is present, the hologram distribution will contain no interference, and can be approximated by a constant. The reconstruction of this hologram can be achieved by analytically calculating the integral in Eq. (37) for $H(X,Y)=1$, and the result will be the exit wave $u(x,y) \approx \frac{1}{z}\exp\left[\frac{i\pi}{\lambda z}(x^2+y^2)\right]$, which is a product of the transmission function $t(x,y)=1$ and the probing divergent spherical wave. Thus, the reconstructed transmission function can be obtained by the division of the reconstructed exit wave by the distribution of the probing wave: $t(x,y) = u(x,y) \Big/ \left\{ \frac{1}{z} \exp\left[\frac{i\pi}{\lambda z}(x^2+y^2)\right] \right\}$.

**Note on plane waves.** Similar considerations can be applied to a plane probing wave. In this case, the interference pattern (hologram) is acquired at some relatively short distance from the sample. The complex-valued object wave at the detector plane is given by the Fresnel diffraction integral, which in the paraxial approximation can be also calculated as a convolution:

$$U(X,Y) = -\frac{i}{\lambda} \iint_S o(x,y) \, \exp\left\{\frac{i\pi}{\lambda z}\left[(x-X)^2 + (y-Y)^2\right]\right\} dxdy, \qquad (42)$$

and the resulting integral transformation is identical to that given in Eq. (37). Hence, for a relatively thin sample which can be described by a 2D transmission function, a hologram recorded with a spherical wave with a source-to-sample distance $z$ has the same distribution as a hologram recorded with a plane wave with a sample-to-detector distance $z$; the only difference is that the hologram distribution is scaled by a magnification factor $M$. Similar considerations also apply to reconstruction. Thus, the reconstruction of each type of hologram, i.e. acquired with plane or spherical waves, is achieved by deconvolution. Algorithms that can be used for the simulation and reconstruction of in-line holograms are discussed in detail in reference [121].

### 3.2.3 Quantitative reconstruction of absorption and phase of objects

Although in-inline holography does not require any optical elements between the sample and the detector, and may appear to be the easiest choice for an experiment, the reconstruction of an in-line hologram is not a trivial task since part of the reference wave is scattered by the object, thus creating an object wave. The reference wave is therefore not well defined. Unlike in off-axis holography, where the reference and object waves are separated, the reconstruction of the complex-valued transmission function of the object from its in-line hologram is not straightforward. It is therefore often stated that phase objects cannot be successfully imaged by in-line holography, but this is not the case, as we will show below.

The following trick allows for the accurate reconstruction of the transmission function of the sample. The transmission function can be written as $t(x,y) = 1 + o(x,y)$, as given by Eq. (32). The sample is illuminated by a incident wavefront $Au_0(x,y)$, where *A* is a complex-valued constant. The reference wave $Au_0(x,y)$ can have an arbitrary distribution, for example a plane wave or a spherical wave. The exit wave is given by $Au_0(x,y)t(x,y)$. Without an object, the wavefront in the detector plane is given by

$$Au_0(x,y) \to AR(X,Y), \qquad |AR(X,Y)|^2 = B(X,Y), \qquad (43)$$

where $B(X,Y)$ is the background. With an object, the wavefront in the detector plane is given by

$$Au_0(x,y)t(x,y) = Au_0(x,y)[1+o(x,y)] \to A[R(X,Y)+O(X,Y)]. \qquad (44)$$

Here, the symbol $\to$ means the forward propagation of the wavefront to the screen plane which is described by the integrals based on the Huygens-Fresnel principle, as provided by Eqs. (9) and (24) (and Eq. (27) in the paraxial approximation). $R(X,Y)=1$ and $R(X,Y)=\exp(ik\rho)/\rho$ for a plane wave and a spherical wave, respectively. The interference pattern on the screen is recorded using a sensitive medium, yielding a hologram with the distribution

$$H(X,Y) = |A|^2 |R(X,Y)+O(X,Y)|^2. \qquad (45)$$

Dividing the hologram distribution by the background distribution gives

$$\frac{H(X,Y)}{B(X,Y)} = |R(X,Y)+O(X,Y)|^2 \approx 1 + R^*(X,Y)O(X,Y) + R(X,Y)O^*(X,Y), \qquad (46)$$

which is the normalised hologram. The resulting normalised hologram does not depend on experimental factors such as the transmission of the medium supporting the sample, the source intensity or the detector sensitivity. By removing these factors, the normalised hologram can be reconstructed using a numerical routine that is independent of the details of the experimental conditions. From the reconstructed complex-valued function $o(x,y)$, the complete transmission function $t(x,y)$ can be obtained by applying Eq. (32), while the absorption and phase distributions of the sample can be reconstructed from $t(x,y)$ by applying Eq. (23). The transmission function of an object $t(x,y)$ reconstructed from the object's hologram can in general be complex-valued, and the correct absorption and phase distributions of the sample can be retrieved in this way [120, 122].

### 3.2.4 The twin image problem

When an in-line hologram is reconstructed, together with the reconstructed object, a twin image appears, as shown in Fig. 10. In the object plane, the twin image appears as a superimposed hologram of the twin object, while in the twin-image plane, the hologram of the object appears superimposed. The two images are therefore inseparable, and are mirror symmetric with respect to the point source. This twin image was the main problem that Dennis Gabor identified in the holography technique he had discovered. Gabor carried out a detailed analysis of the origin of this twin image [20], but did not offer a solution for separating the reconstructed object from its twin.

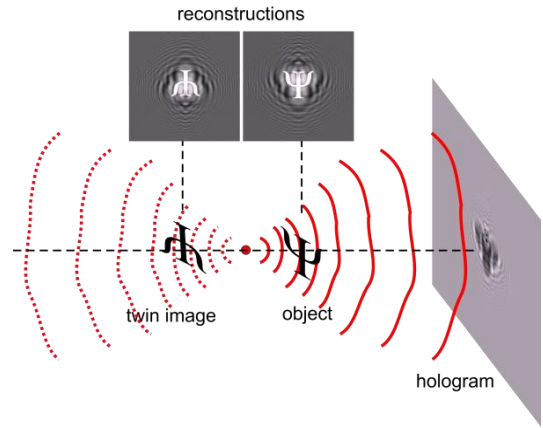

Fig. 10. Formation of the twin image during reconstruction of an in-line Gabor hologram. Both images are symmetrical relative to the point source. Adapted from [123].

In an attempt to solve the twin image problem, optical off-axis holography was invented in 1962 by Leith and Upatnieks [124]. This approach separated the directions of the object and reference waves, but required re-arrangement of the optical scheme. It should be also mentioned that the first electron off-axis holograms had previously been demonstrated in 1956–1957, shortly after the invention of the biprism, which splits an electron beam into two parts [125, 126]. In off-axis holography, the object and twin image do not share the same optical axis, and are physically separated. Off-axis optical holography became the most popular type of holography for artistic purposes, such as holographic portraits. However, the problem of the twin image in *in-line* Gabor holography remained.

A numerical solution to the twin image problem was presented in 2007 [123]. The aforementioned possibility of reconstructing the absorption and phase properties of an object allows for a careful inspection of the reconstructed absorption distribution. When an in-line hologram is reconstructed, the interference between the reconstructed object wavefront and the twin image wavefront creates negative values in the reconstructed absorption distribution. Since the physical values of the absorption cannot be negative, these negative values can be attributed to the presence of the twin image. A constraint of non-negative absorption can then be imposed, in which all negative absorption values are set to zero and positive absorption values are left unchanged. When implemented in an iterative reconstruction routine, this constraint allows for the suppression of the negative absorption values and the interference due to the presence of the twin image can be suppressed. In this manner, the twin image can be removed and the true object distribution can be retrieved, with its absorption and phase-shifting distributions [123]. This iterative method of twin-image-free reconstruction does not depend on the wavelength of the waves employed, and has been successfully applied in light optical [123, 127], electron [128], and terahertz [35, 129] in-line holography.

A general scheme of the iterative phase retrieval and reconstruction of an in-line hologram is shown in Fig. 11. Here, the sample must be sufficiently thin that it can be approximated by a 2D distribution in a one plane (for example, polystyrene spheres on glass). The constraints are then applied in two planes: the sample plane and the hologram plane [123].

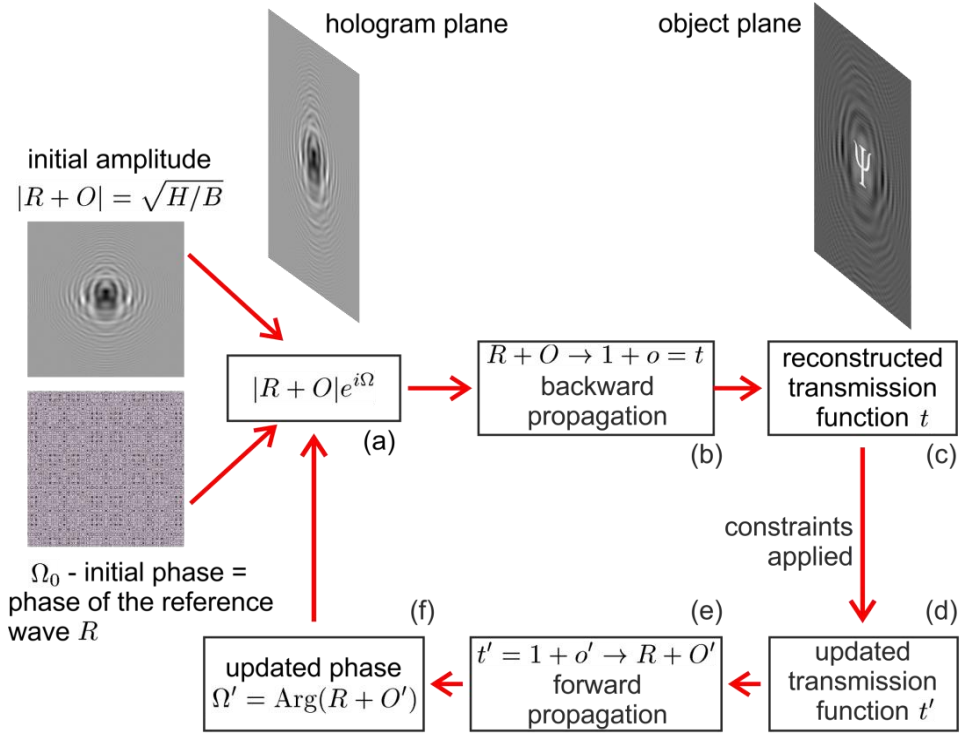

Fig. 11. A general scheme for iterative phase retrieval from a single-shot intensity measurement (hologram). (a) The algorithm starts in the hologram plane, where the initial complex-valued distribution is created by combining the measured amplitude distribution with the phase of the reference wave. (b) The wavefront propagates from the hologram plane to the sample plane, where it gives the distribution of the complex-valued transmission function $t(x, y)$. (c) Constraints are imposed in the sample plane. (d) The updated sample transmission function $t'(x, y)$ is obtained. (e) The wavefront is then propagated from the sample plane to the detector plane (f). The amplitude of the wavefront distribution in the hologram plane is replaced with the measured amplitude, and thus the complex-valued wavefront distribution in the detector plane is updated for the next iteration in (a). Adapted from [123].

### 3.2.5 Reconstruction of the phase object from a single-shot in-line hologram

In general, a phase object cannot be reliably recovered from a single-shot in-line hologram without applying an iterative reconstruction method, as demonstrated in reference [130]. Moreover, reconstruction by simple backward wavefront propagation (Eq. (36)) will give misleading results, as illustrated in Fig. 12. Figure 12 which shows a simulated in-line hologram of an object with absorbing and phase properties and its reconstructions. The amplitude and the phase of the transmission function in the object plane vary in the ranges 0.6–1 au and 0–2 rad, respectively, as shown in Fig. 12(a). The phase of the transmitted wave in the detector plane reaches 1 rad, as shown in Fig. 12(b); this phase distribution is lost during intensity measurement. The amplitude and the phase distributions retrieved by non-iterative reconstruction are shown in Fig. 12(c). Both distributions exhibit superposition of the reconstructed object distribution and its twin image, which appears as a distribution of concentric rings rather than as a well-defined object. From these reconstructions, it is not evident that the object is reconstructed at the correct in-focus position, and this can be a problem when reconstructing an experimental hologram, where the exact in-focus position of the object is unknown. Moreover, neither the amplitude nor phase distributions are reconstructed quantitatively correctly. Fig. 12(d) shows the amplitude and the phase distributions reconstructed by applying an iterative phase retrieval procedure, as described in reference [130]; both distributions are almost perfectly recovered.

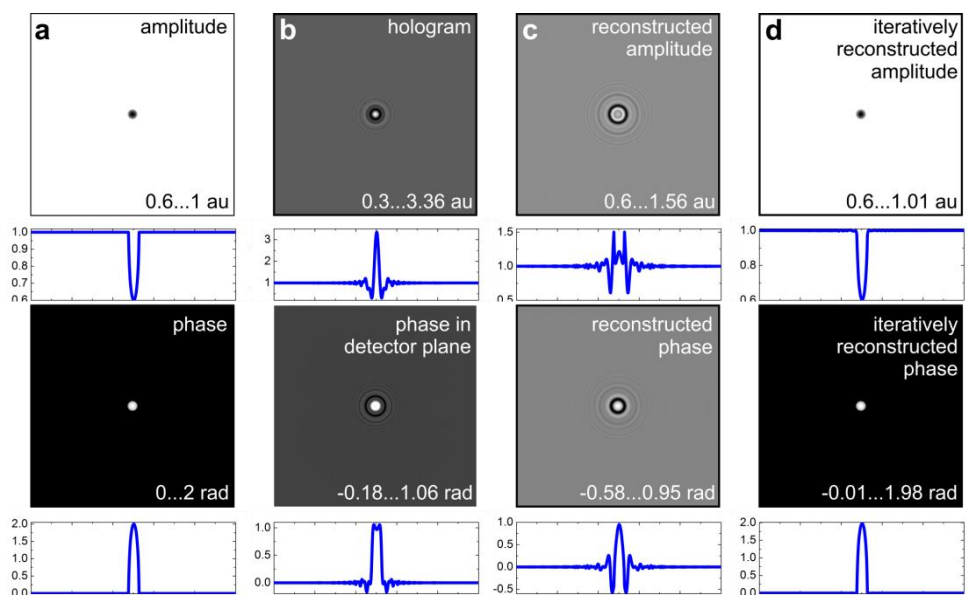

Fig. 12. Simulated in-line hologram of an object with absorbing and phase properties, and its reconstructions. (a) Distributions of the amplitude (top) and phase (bottom) of the transmission function of the object. (b) Simulated hologram (top) and phase distributions at the detector plane (bottom). (c) Reconstructed amplitude (top) and phase (bottom) distributions of the transmission function (reconstruction is carried out

via backward wavefront propagation, as described in Eq. (36)). (d) Iteratively reconstructed amplitude (top) and phase (bottom) distributions of the transmission function in the object plane. The blue curves show the profiles through the middle of the corresponding images. Adapted from [130].

An example of an iteratively reconstructed phase object from its experimental in-line electron hologram is shown in Fig. 13. In this case, an iterative reconstruction procedure similar to that shown in Fig. 11 was applied, with an additional constraint in the object plane, a tight support constraint, as indicated by the blue lines in Fig. 13(a). In the resulting reconstructions, the twin image is removed and the amplitude and the phase distributions are recovered to their correct values (Fig. 13(b) and (c)).

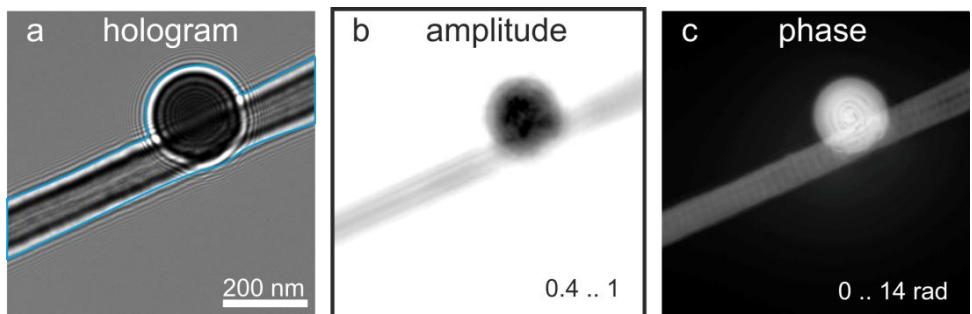

Fig. 13. Electron (200 keV) in-line hologram of a latex sphere and its reconstruction. (a) In-line hologram of the sphere, acquired under defocus (180 μm). The blue lines indicate the support, outside of which the transmission was set to one during the iterative reconstruction. (b) Reconstructed amplitude distribution. (c) Reconstructed phase distribution. Reprinted from [128], with permission from Elsevier.

### 3.2.6 3D sample reconstruction from two or more in-line holograms

The absorption and the phase distributions of a thin 2D sample can be reconstructed from a single-shot in-line hologram, as explained in the previous sub-section. However, realistic samples are described by rather 3D than 2D distributions. In optical holography, a complete wavefront reconstruction from a sequence of intensity measurements by applying an iterative reconstruction procedure was reported in series of studies between 2003 and 2006 [131-135]. In electron holography, a similar approach is known as focal series reconstruction [136], which is applied to obtain unambiguous, high-resolution reconstruction of samples from a focal series acquired via high-resolution transmission electron microscopy (HRTEM) [137]. It has recently been demonstrated that 3D samples, including 3D phase objects, can be reconstructed from two or more in-line holograms acquired at different *z*-distances from the sample [122] (Fig. 14(a)). In this method, the reconstruction is achieved via iterative wavefront propagation between the planes in which the intensity distributions are measured ($H_1$ and $H_2$ in Fig. 14(a)), without involving any planes within the sample and hence without requiring constraints on the sample. Once the compete complex-valued wavefront has been recovered, it can be then propagated backwards to the sample planes, thus reconstructing the 3D distribution of the sample (Fig. 14(b)). In principle, as few as two holograms are sufficient to reconstruct the entire wavefront diffracted by a 3D sample. There is no restriction on the thickness or sparsity of the sample, since a reference wave is not required. This method can be applied to 3D samples such as 3D distributions of particles, dense biological samples, 3D distributed phase objects, etc.

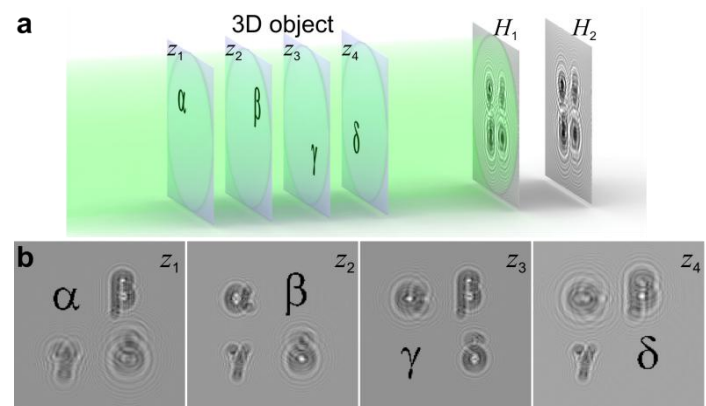

Fig. 14. Reconstruction of a 3D distributed sample from two or more intensity measurements. (a) Experimental arrangement, in which the 3D sample is represented by a set of planes at different $z$-positions and two holograms are acquired at different distances from the sample, $H_1$ and $H_2$. (b) Reconstructed amplitude distributions at the four planes within the 3D sample distribution. Adapted from [122].

### 3.3 Gabor in-line holography with low-energy electrons

#### 3.3.1 Experimental examples
The in-line holographic imaging of macromolecules with low-energy electron microscopes, as carried out by Professor Fink's group at the University of Zurich (shown in Fig. 9), has been reported in the literature [17, 37, 38, 41-43, 138-141], and some examples are shown in Fig. 15. These reconstructions were obtained by solving the integral transform given in Eqs. (35)–(37), as described in [121]. Despite the very short wavelength (0.8–1.7 Å) of the probing wave, the resolution of the reconstructed structures is about 1 nm, and the atomic details of molecules remain unresolved.

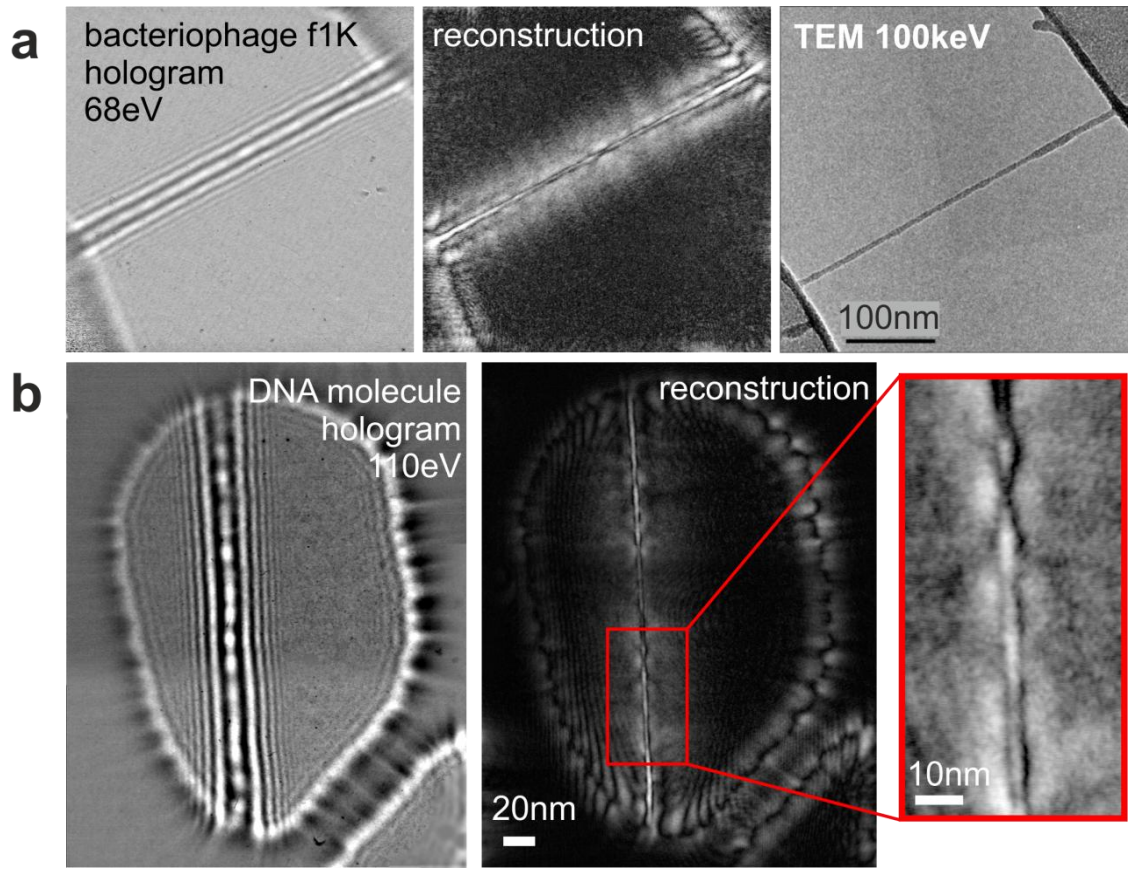

Fig. 15. Low-energy electron in-line holography imaging of biomolecules. (a) Individual bacteriophage [41], reprinted by permission from Springer Nature, copyright 2011. (b) Individual DNA molecules [37], copyright Springer Nature 2013.

### 3.3.2 Biprism effect

One challenge associated with the use of low-energy electrons in imaging is the high sensitivity of low-energy electrons to the local inhomogeneous potential distribution of the sample. Previous studies have demonstrated that when a freestanding object is positioned close to the source of coherent low-energy electrons, the trajectories of the electrons are deflected towards the object. This effect is similar to the deflection of electrons by a positively charged wire [125], and is therefore called the biprism effect [39], illustrated in Fig. 16(a) and (b). This effect affects the entire electron wave, and as a result, the reference wave is distorted, which complicates both the reconstruction procedure and the interpretation of the retrieved structures. An example is shown in Fig. 16(c) and (d), where the sample distribution reconstructed from the in-line hologram does not show a well-defined object. As a solution to this problem, the reconstruction step can be replaced by the simulation of holograms of positively charged objects that would fit the experimental data [29, 142, 143].

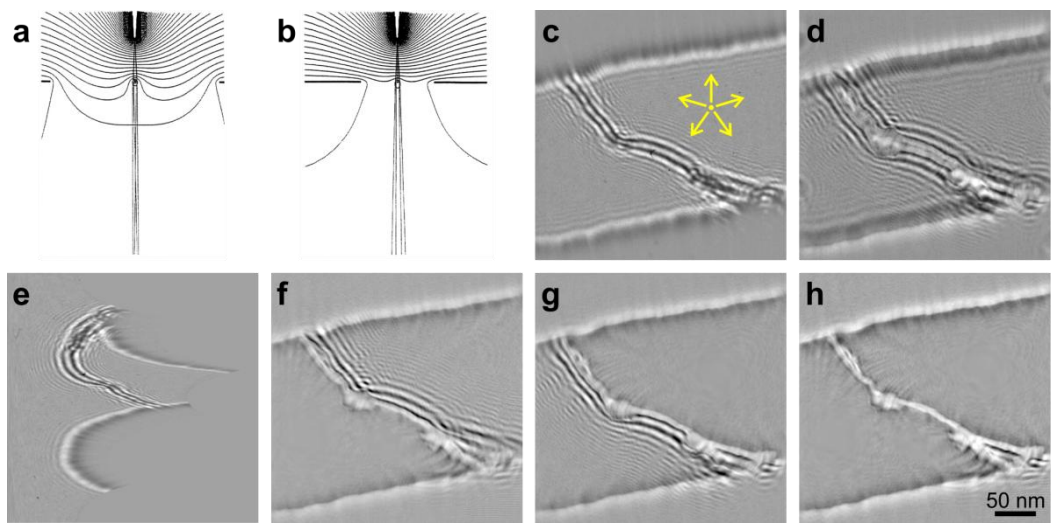

Fig. 16. Biprism effect and single-sideband holography reconstruction of an experimental low-energy electron hologram. (a) Ray tracing results for field emission tip above a fiber stretched across a hole in a support film. Tip located above the center of the hole. The support film and the fiber are at ground potential, the tip at –40 V; fiber diameter, 18 nm; hole diameter, 2 μm (same as Quantifoil hole). Rays shown have emission half angle of 4° and 6°, bending of inner rays 3°. (b) Same as (a) except that the tip is located 900 nm off the center of the hole (100 nm away from the edge of the hole). Bending of rays is reduced. (a) and (b) reprinted from [39], with permission from Elsevier. (c) Hologram of a bundle of multi-walled carbon nanotubes acquired with 100 eV electrons (the yellow dot and arrows indicate the position of the centre and the directions of transformation to polar coordinates for further sideband filtering). (d) Distribution reconstructed from the hologram by conventional reconstruction procedure, demonstrating failure to correctly retrieve the molecules shape. (e) Hologram transformed into polar coordinates. (f) Reconstruction obtained from the left sideband filtered hologram. (g) Reconstruction obtained from the right sideband filtered hologram. (h) Final reconstructed distribution obtained by combining the left and right sideband filtered reconstructions. (c)–(h) Reprinted from [139] with permission from Elsevier.

The biprism effect is not always disruptive, as demonstrated by the successful reconstructions in Fig. 15; however, it becomes more pronounced at shorter source-to-sample distances, where higher magnification and thus higher resolution is expected. A solution to this problem was proposed in [139], which relies on single-sideband holography [144]. In this approach, half of the Fourier spectrum of the hologram is set to zero before the reconstruction, which eliminates the twin image on the corresponding side. For an arbitrarily shaped object, a coordinate transformation is applied before Fourier filtering, which ensures that the object distribution is located on one side. After Fourier filtering, the resulting reconstruction is biprism- and twin-image free [139], as illustrated in Fig. 16(e)–(h).

An experimental solution to the biprism problem uses graphene as a sample support [145]. Graphene is conductive, and provides a nearly flat grounded plane for electrons, which helps in avoiding distortion due to the high field gradients that form around suspended biased nanoscale objects [146]. In the next subsection, we consider the imaging of graphene with low-energy electrons in transmission mode.

### 3.3.3 Low-energy electron in-line holography of graphene

Graphene has been successfully used as a sample support [145] for the low-energy electron in-line holographic imaging of gold nanorods [140], biological macromolecules [40, 43] and alkali metal clusters [147]. Moreover, graphene itself exhibits various interesting phenomena when studied with low-energy electrons. The high sensitivity of low-energy electrons to local electric potentials allows for the detection of charged adsorbates on a graphene surface with a sensitivity of a fraction of an elementary charge [148], as shown in Fig. 17. A positively charged adsorbate (Fig. 17(a)) acts as a tiny lens, focusing electrons to a bright spot, Fig. 17(b) shows several such bright spots in an experimental hologram. Low-energy electrons have a sensitivity to local potentials that is hundreds times higher than that of high-energy electrons. This can be roughly explained by the fact that slower electrons spend more time in the potential and thus are more strongly deflected.

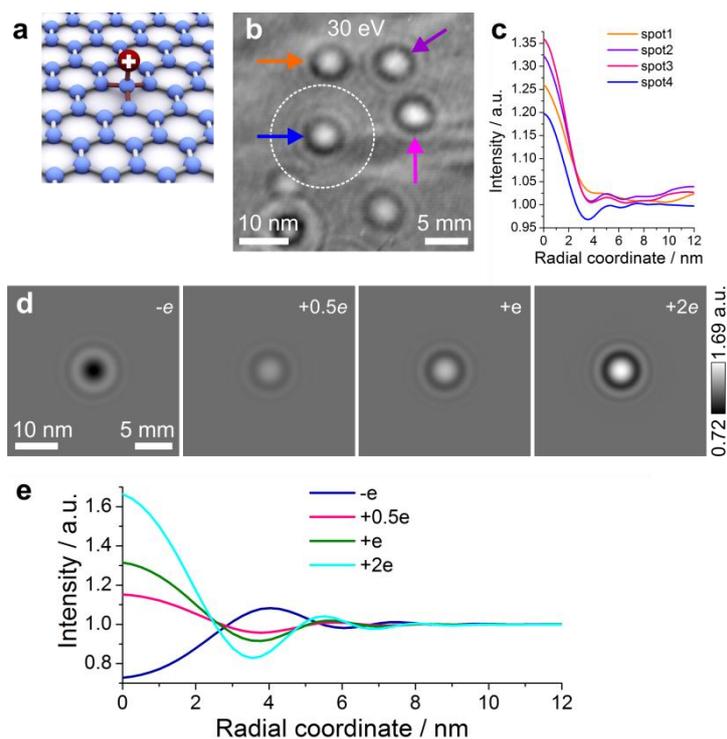

Fig. 17. Low-energy electron in-line holograms of charged adsorbates. (a) Schematic representation of a charged adsorbate on graphene. (b) Experimental hologram exhibiting bright spots (electron energy is 30 eV, source-to-sample distance is 82 nm and source-to screen distance is 47 mm). (c) Angular-averaged intensity profiles of the four bright spots shown in (b). (d) Simulated in-line holograms of a point charge at four different charge values, where the simulation parameters match those of the experimental hologram shown in (b). (e) Angular-averaged intensity profiles as a function of the radial coordinate, calculated from the simulated holograms shown in (d). The scale bars in (b) and (d) indicate the sizes in the object plane (left) and in the detector plane (right). Adapted with permission from [148], copyright (2016) American Chemical Society.

The iterative reconstruction of low-energy electron in-line holograms of individual charges gives the object amplitude and phase distributions associated with the absorption and potential distributions, respectively [149] (Fig. 18). The reconstructed absorption distributions (Fig. 18 (c) and (d)) appear to be more localised than the reconstructed phase distributions (Fig. 18 (e) and (f)). The absorption distribution is related to the actual size of the adsorbate, while the phase distribution is related to the potential created by the adsorbate's charge.

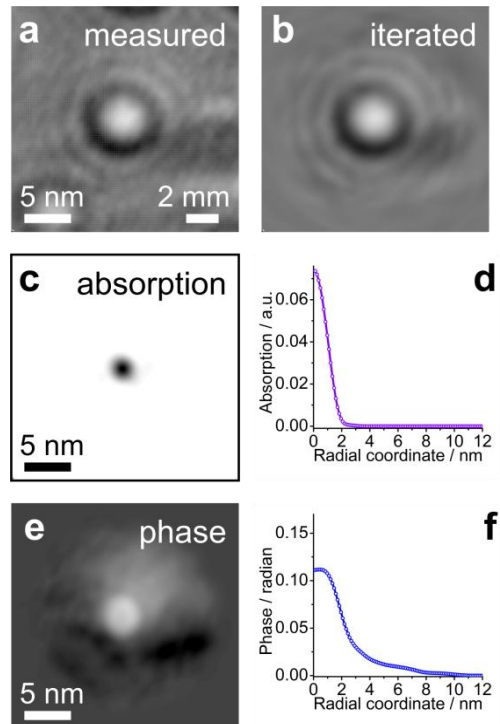

Fig. 18. Iteratively reconstructed absorption and phase distribution of an individual adsorbate. (a) In-line hologram recorded with 30 eV electrons, showing a bright spot. (b) Intensity distribution obtained after 2000 iterations, which matches the experimental distribution. (c) and (d) reconstructed absorption and phase distributions, respectively. (e) and (f) corresponding angular-averaged profiles. Reprinted from [149], with permission from Elsevier.

### 3.3.4 Mapping unoccupied electronic states of 2D materials

Mapping unoccupied electronic states of 2D material such as graphene can be realized by employing low-energy electron in-line holography experimental setup at very short source-to-sample distances and at very low associated extracting voltages. The images acquired as a function of the energy of the probing electron beam allow measurement of energy- and angle-resolved low-energy electron transmission, as was demonstrated in [150] and illustrated in Fig. 19. The technique can be applied to study various 2D materials provided that they are sufficiently transparent to low-energy electrons and can be prepared freestanding.

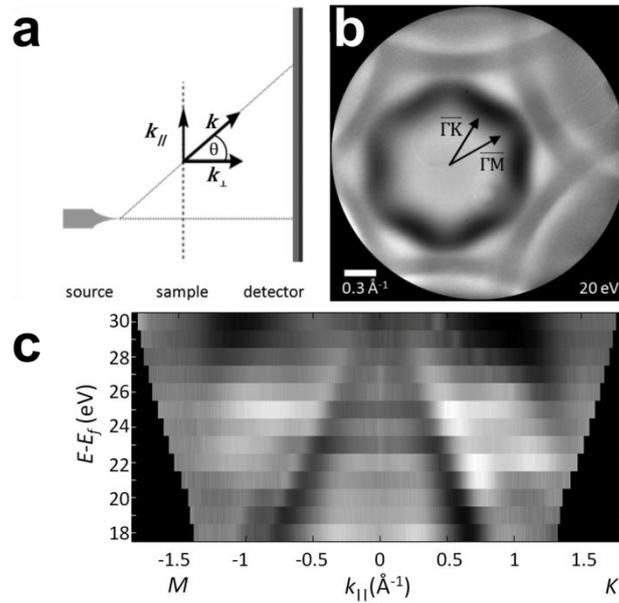

Fig. 19. Mapping unoccupied electronic states of freestanding graphene. (a) Schematics illustrating the determination of the in-plane momentum components of the incident electrons. Electrons are field-emitted from a sharp tip and arrive at the freestanding graphene sheet under a large angular range. The in-plane momentum of an electron in the graphene plane is related to the probing angle $\theta$ and the total momentum $\hbar k = \sqrt{2m_e E_{\text{kin}}}$. While the kinetic energy is determined by the tip bias potential and the work function of the sample, the angle can be determined from the position where the electron is recorded in the detector plane. (b) Background corrected transmission intensity recorded with an electron energy of 20 eV. (c) Plot of the background corrected transmission intensity as a function of the electron energy and the $k_{\parallel}$ values in the $\overline{\Gamma M}$ and $\overline{\Gamma K}$ direction of the Brillouin zone. Figure reprinted from [150], copyright (2016) by the American Physical Society.

### 3.4 CDI with low-energy electrons

Low-energy electron diffraction patterns of various individual macromolecules, for example carbon nanotubes [117, 138], and graphene [118, 151] were acquired in a dedicated low-energy electron microscope equipped with a microlens [116] to collimate the electron beam, as shown in Fig. 9(b). Diffraction patterns of individual stretched single-walled carbon nanotubes (SWCNTs) were acquired at a resolution of 1.5 nm [117], as shown in Fig. 20(a)–(c). Diffraction patterns of graphene and bilayer graphene are shown in Fig. 20(d)–(f). Diffraction patterns of bundles of individual carbon nanotubes were reconstructed using a holographic CDI (HCDI) approach at a resolution of 0.7 nm [138] and are shown in Fig. 24.

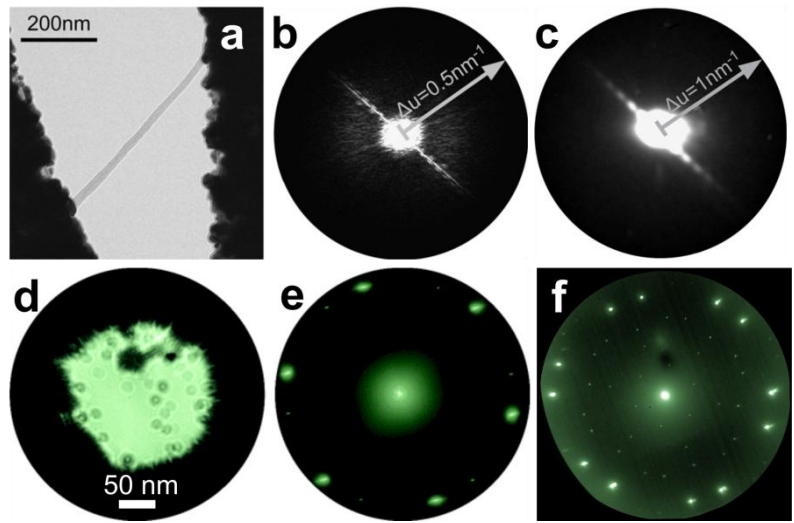

Fig. 20. Coherent diffraction imaging (CDI) with low-energy electrons. (a)–(c) CDI of an individual single-walled carbon nanotube (SWCNT) (reprinted from [117], with permission from Elsevier). (a) Transmission electron microscopy (TEM) image of the sample. (b) Fourier transform of the TEM image. (c) Diffraction pattern of SWCNTs recorded with 186 eV electrons. (d)–(e) In-line holography and CDI of graphene [118]. (d) In-line hologram of graphene sample recorded using electrons with kinetic energy 58 eV, source-to-sample distance 380 nm, and source-to-detector distance 68 mm. (e) Diffraction pattern of the same sample area as in (e) recorded using electrons with kinetic energy 236 eV. (f) Diffraction pattern of bilayer graphene recorded using electrons with kinetic energy 236 eV and source-to-detector distance 68 mm; here, besides the intense first-order peaks, numerous weaker moiré peaks are observed [151] (reprinted from [151], with permission from Elsevier).

## 4. Future directions

### 4.1 Volumetric 3D deconvolution

The beauty of holography is that it offers the possibility of restoring 3D objects from their 2D holograms, as often demonstrated in optical art holography. In digital holography, however, when a 2D hologram is numerically reconstructed instead of a 3D object, the object wavefront is reconstructed and further analysis is required to extract the 3D object itself. The reconstruction of digital holograms is achieved via the calculation of integral transformations, as described above (Eqs. (35)–(37)). As a result, the 3D object wavefront is obtained as a set of 2D complex-valued distributions at various distances from the hologram plane (or, at different cross-sections of the sample distribution), in an analogous way to 'optical sectioning' in optical microscopy. In every such 2D slice, there is both a signal from the in-focus part of the object and a blurred signal from the out-of-focus part of the object. Thus, the reconstructed object wavefront does not directly represent the original 3D object itself, as illustrated in Fig. 21. The problem is then to extract the 3D object distribution from its reconstructed scattered wave distribution.

A 3D volumetric deconvolution method has been proposed to retrieve the original 3D distribution of the object from its reconstructed wavefront [152]. In this approach, the 3D complex-valued optical wavefront scattered by an object $u_\text{O}(\vec{r})$ can be represented as a convolution of the 3D object distribution $o(\vec{r})$ and the 3D complex-valued point-spread function $u_\text{PSF}(\vec{r})$:

$$u_\text{O}(\vec{r}) = o(\vec{r}) \otimes u_\text{PSF}(\vec{r}), \tag{47}$$

where $u_\text{PSF}(\vec{r})$ is the wavefront scattered by a point scatterer, and $\vec{r}$ is a 3D coordinate. $u_\text{O}(\vec{r})$ is the wavefront reconstructed from the hologram of the object, while $u_\text{PSF}(\vec{r})$ is the wavefront reconstructed from the hologram of a point scatterer, recorded under exactly the same condition as the hologram of the sample, where the position of the point scatterer may be approximately in the centre of the sample distribution. The hologram of the point scatterer can be recorded experimentally or simulated, with the latter being a much simpler approach. A simulated point spread function (PSF) is preferred because the point scatterer can be designed as a truly point-like object, unlike in an experiment. Provided both distributions $u_\text{O}(\vec{r})$ and $u_\text{PSF}(\vec{r})$ are available, $o(\vec{r})$ can be found by solving Eq. (47), which requires a 3D deconvolution of $u_\text{O}(\vec{r})$ with $u_\text{PSF}(\vec{r})$. The principle of 3D deconvolution is illustrated in Fig. 21, and should not be confused with conventional 2D deconvolution, which is applied in digital holography to achieve reconstruction. In digital holography, "reconstruction by deconvolution" usually refers to the 2D deconvolution of a 2D hologram with a free-space propagator, as expressed above in Eqs. (35)–(37). This is a procedure that is always needed to obtain the initial reconstructions of the complex-valued reconstructed object wave. What follows next, however, is a completely different type of deconvolution: a 3D deconvolution applied to the reconstructed 3D complex-valued fields that allows us to obtain the object itself, rather than just the object wave as obtained in a traditional reconstruction.

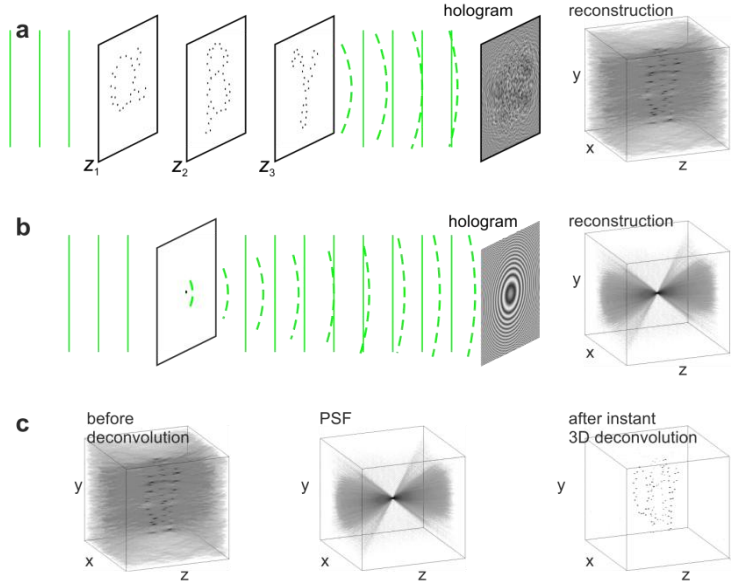

Fig. 21. Principle of 3D deconvolution in digital holography. (a) Recording a hologram of a 3D sample and the 3D distribution of the amplitude of the complex-valued wavefront reconstructed from the hologram. (b) Recording a hologram of a point scatterer and the 3D distribution of the amplitude of the complex-valued 3D point spread function (PSF) obtained by reconstruction of the hologram. (c) The 3D deconvolution of the 3D reconstructed wavefront (a) with the 3D PSF (b) gives the 3D reconstruction of the 3D sample.

Two methods of 3D deconvolution have been proposed: instant deconvolution and an iterative deconvolution. Instant 3D deconvolution is achieved by applying the formula:

$$o'(\vec{r}) = \mathrm{FT}^{-1}\left( \frac{\mathrm{FT}\left(|u_\mathrm{O}(\vec{r})|^2\right)}{\mathrm{FT}\left(|u_\mathrm{PSF}(\vec{r})|^2 + \beta\right)} \right), \quad (48)$$

where $|u_\mathrm{O}(\vec{r})|^2$ is the 3D reconstructed intensity of the reconstructed object wave, $|u_\mathrm{PSF}(\vec{r})|^2$ is the 3D reconstructed intensity of a point scatterer, which is the PSF of the system, $\mathrm{PSF}(\vec{r})$; and $\beta$ is a small addendum to avoid division by zero. The 3D instant deconvolution given in Eq. (48) is equivalent to the Wiener filter in signal processing. The resulting function $o'(\vec{r})$ is not the distribution of the sample, but it is the distribution of the exact *positions* of the individual scatterers. This method can be applied effectively to the distribution of point-like objects, such as particles in a solution.

The results of applying 3D volumetric deconvolution to experimental optical holograms are shown in Fig. 22. While conventional reconstruction results in an out-of-focus signal superimposed at the object position (Fig. 22(b)), reconstruction after 3D volumetric deconvolution is completely free of the out-of focus signal (Fig. 22(c)), and the positions of the individual scatterers are resolved. Using this method, the out-of focus signal and twin image are removed, and spatially well-localised parts of the sample are recovered. 3D deconvolution also improves the lateral resolution of the reconstructed object distribution, as shown in Fig. 22(d)–(e) [152].

3D volumetric deconvolution by instant deconvolution has successfully been applied to 3D particle field reconstruction and particle tracking [153, 154], and 3D iterative deconvolution has been recently demonstrated using biological samples [155].

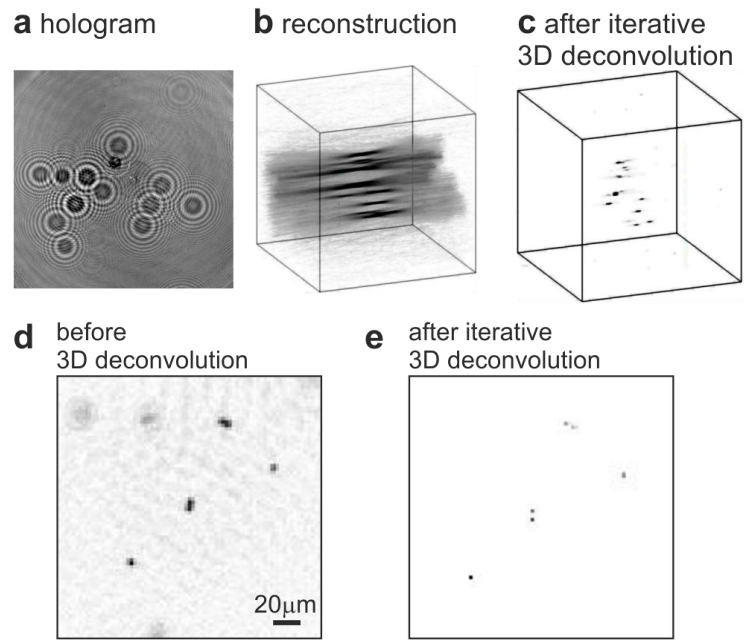

Fig. 22. 3D volumetric deconvolution of experimental optical holograms. (a) Normalised optical hologram of polystyrene microspheres of diameter 4 µm, placed on both sides (S1 and S2) of microscopic glass 170 µm thick. (b) 3D representation of the reconstructed amplitude distribution. (c) 3D representation of the results after 3D volumetric iterative deconvolution. (d) and (e) magnified images of the reconstructed amplitude in a selected plane (S1) before and after iterative 3D volumetric iterative deconvolution, showing the improvement in the lateral resolution. Adapted from [152].

## 4.2 Merging holography and coherent diffractive imaging, HCDI

To overcome the shortcomings of CDI, which are mainly related to the ambiguity of the reconstruction, the novel experimental techniques described above have been used to merge CDI with the holographic approach, thus capturing the phase distribution of the scattered object wave (for example in FTH [74-76, 79-86] and Fresnel CDI [87-90]), but these techniques require advanced experimental arrangements. In fact, there is a direct relationship between holography and CDI [138]: the FT of an in-line hologram distribution $H(X,Y)$ is proportional to the complex-valued distribution of the scattered object wave in the far field, $\text{FT}[t(x,y)]$. This relationship can be proven as follows. In CDI, the measured intensity in the far field is given by:

$$I(u,v) = \left|\text{FT}[t(x,y)]\right|^2. \tag{49}$$

In in-line holography, reconstruction of an object transmission function $t(x,y)$ from its in-line hologram $H(X,Y)$ requires deconvolution with the Fresnel function, according to Eq. (37) [138]:

$$\text{FT}\left\{\exp\left[-\frac{i\pi z}{\lambda Z^2}(X^2+Y^2)\right]\right\}\text{FT}[H(X,Y)] \approx \text{FT}[t(x,y)]. \tag{50}$$

Eqs. (49) and (50) demonstrate that the modulus of the FT of the hologram $\left|\text{FT}[H(X,Y)]\right|$ gives the amplitude of the scattered wave in the far field $\left|\text{FT}[t(x,y)]\right|$. Moreover, since the phase of $\text{FT}\left\{\exp\left[-\frac{i\pi z}{\lambda Z^2}(X^2+Y^2)\right]\right\}$ is known, the phase of the scattered wave in the far field, $\text{FT}[t(x,y)]$, can be determined from the phase of the FT of the hologram, $\text{FT}[H(X,Y)]$. Thus, a good estimate of the phase in the far field can be obtained from the FT of the hologram, and can be plugged into the first iteration of the iterative phase retrieval routine, where a random phase distribution is conventionally employed. In this case, the iterative phase retrieval quickly converges to a stable solution. For comparison, the FT of the real-space image of the sample distribution will not provide the correct distribution for the phase in the far field, since the real-space image does not contain the phase distribution of the sample. In addition, the diffraction pattern and hologram can be acquired in a single experimental scheme by transforming the probing wave from a plane to a spherical wavefront.

Holographic CDI (HCDI) requires two experimental images of an object: a hologram and a diffraction pattern. The FT of the hologram provides the phase distribution and hence the solution to the 'phase problem' in a single step. The diffraction pattern is then required to recover the high-resolution information using an iterative routine. In addition, the central region of the diffraction pattern, which is usually missing, can be adapted from the amplitude of $\text{FT}[H(X,Y)]$. The advantage of HCDI over other techniques is that a standard CDI experimental setup can be adapted for HCDI simply by making the wavefront slightly divergent, to allow for in-line hologram recording. For example, in the low-energy electron microscope sketched in Fig. 9, this is achieved by changing the voltage at the collimating microlens. During acquisition of the hologram and diffraction pattern, the sample remains at the same position in the beam, and this is another important advantage over conventional CDI, in which the sample must be imaged by a complementary technique to obtain the low-resolution information required to create the masking support in the phase retrieval routine.

HCDI has been successfully demonstrated in light optical and low-energy electron experiments, as shown in Figs. 23 and 24 [138].

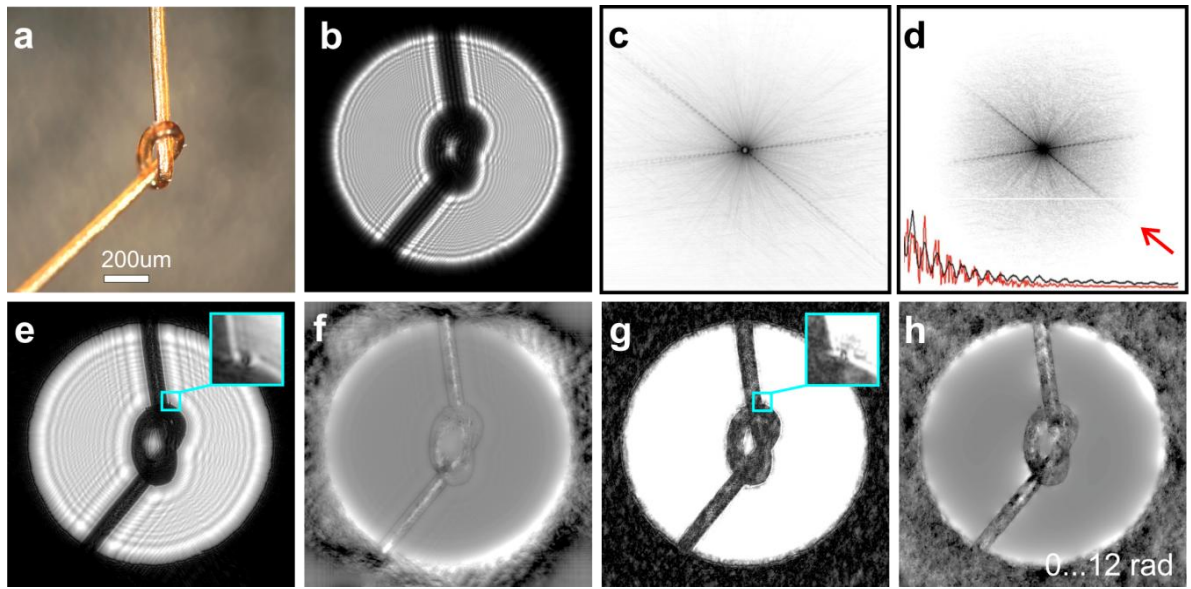

Fig. 23. Optical HCDI of a human hair. (a) Reflected-light microscopy image of a hair. (b) Recorded optical in-line hologram with spherical wave (source-to-sample distance 5.3 mm). (c) Experimental diffraction pattern recorded at 1 m from the sample. (d) Amplitude of the FT of the hologram shown in (b), where the inset shows the profiles of the square root of the experimental diffraction pattern intensity (black) and the amplitude of the FT of the hologram (red) along the direction indicated by the red arrow. The resolution of a hologram can be estimated by the highest visible frequency in the spectrum (the FT of the hologram), and is usually less than the resolution provided in the diffraction pattern. (e) Amplitude distribution of the sample reconstructed from the hologram. The superimposed interference pattern arises from the twin image. A magnified region of the reconstruction is shown in the inset. (f) Phase distribution of the sample reconstructed from the hologram. (g) Sample amplitude distribution reconstructed by HCDI. The inset contains a magnified region of the reconstruction, showing the improved resolution in comparison to the hologram reconstruction. (h) Phase distribution of the sample, reconstructed using HCDI. Adapted from [138].

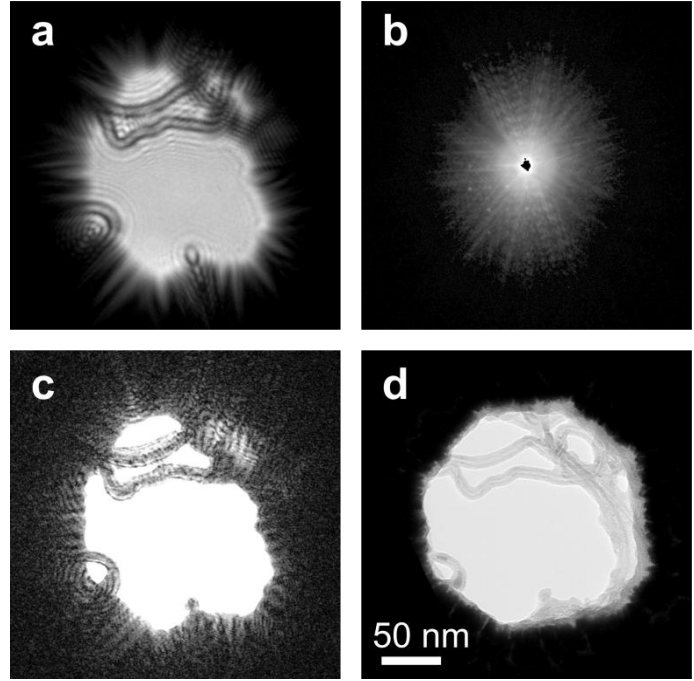

Fig. 24. Low-energy electron HCDI of individual carbon nanotubes, with a distance between the electron source and detector of 68 mm. (a) In-line hologram recorded with electrons of kinetic energy 51 eV and source-to-sample distance 640 nm. (b) Diffraction pattern recorded with electrons of kinetic energy 145 eV, with resolution $R = \lambda/(2NA) = 7$ Å. (c) Reconstructed amplitude distribution using HCDI. (d) TEM image recorded with 80 keV electrons. Adapted from [138].

## 4.3 Extrapolation

The resolution of an optical system is estimated using the Abbe criterion, $R = \lambda/(2NA)$, where $NA$ is the numerical aperture of the optical system [10-12]. Based on this criterion, the sole limit of the resolution in lensless imaging (besides the wavelength) is the size of the interference pattern (detector). It has recently been demonstrated that when coherent waves are used, the recorded interference pattern contains sufficient information to *extrapolate* the recorded pattern beyond the detector area, and thus to effectively increase the resolution *a posteriori* [156-159].

The iterative reconstruction procedure with extrapolation includes steps that are similar to the conventional iterative phase retrieval procedure (Figs. 4 and 11), as shown in Fig. 25. For a hologram (or diffraction pattern) $H_0$ of size $N_0 \times N_0$ pixels, the iterative reconstruction includes the following steps:

(i) Formation of the input complex-valued field in the hologram plane $U(X,Y)$. The amplitude of the central $N_0 \times N_0$ part is always given by the square root of $H_0$, and the amplitude of the remaining pixels outside $H_0$ of the $N \times N$ array is updated at each iteration. In the first iteration, these values are set to a constant value or are randomly distributed. The phase distribution is initially set to the phase of the known reference wave, and is also updated at each iteration.

(ii) The complex-valued wavefront is propagated from the detector to the object plane.

(iii) In the object plane, the complex-valued object transmission function $t(x,y)=1+o(x,y)$ is reconstructed and the following constraints are applied. Since the object has a finite size, the distribution $o(x,y)$ is multiplied with a loose mask that sets the values outside a certain region to zero [128], as shown in Fig. 25. A second constraint is based on the physical notion that the absorption must be positive [123]; consequently, the pixel values where absorption is negative are set to zero. As a result, the transmission function in the object plane is updated to $t'(x,y)=1+o'(x,y)$, and the updated complex-valued exit wave is obtained.

(iv) The updated complex-valued exit wave is propagated from the object plane to the detector plane, thus giving the updated wavefront in the detector plane $U'(X,Y)$. The amplitude and the phase distributions of $U'(X,Y)$ form the input values for the next iteration in step (i).

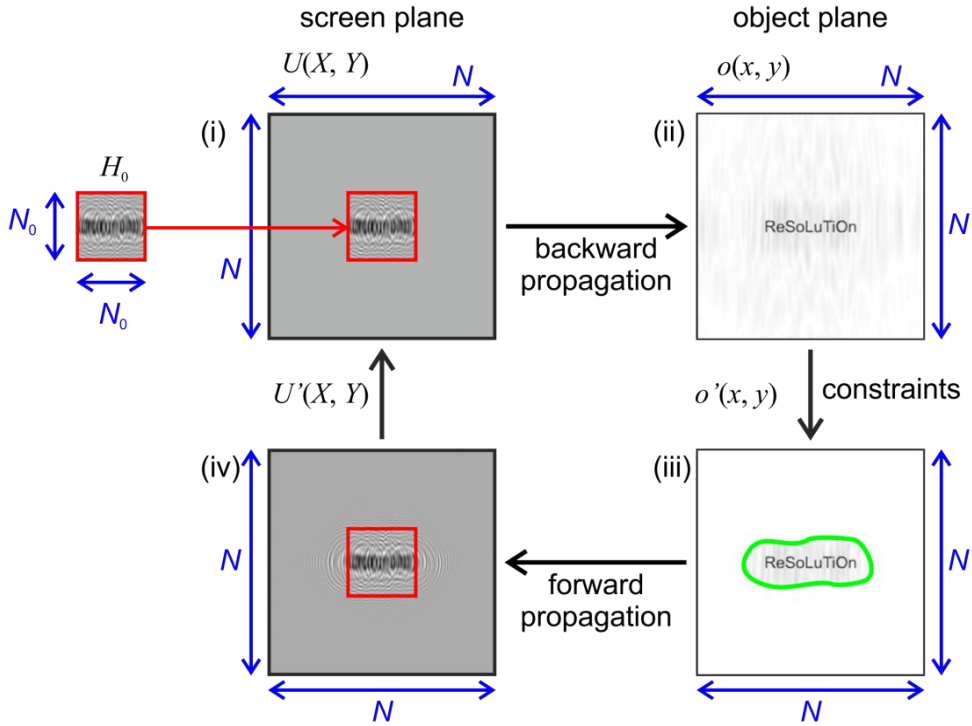

Fig. 25. Iterative self-extrapolation reconstruction of a hologram of an object (the word "ReSoLuTiOn"). The iterative loop uses steps (i)-(iv) as described in the main text. The reconstructed object amplitude distributions in (ii) and (iii) are shown with an inverted intensity scale. Adapted from [156].

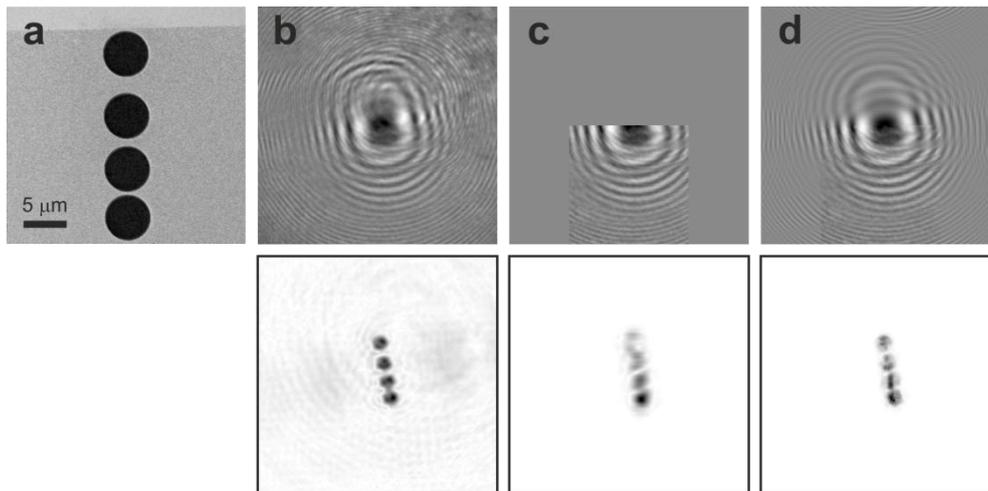

Fig. 26. Resolution enhancement in digital optical holography by extrapolation of a hologram. (a) Scanning electron microscopy image of the sample. (b) 1000 × 1000 pixel experimental optical hologram of the sample (top) and its reconstruction (bottom). (c) Selected 500 × 500 pixel part of the original hologram (b) is padded with zeros to give 1000 × 1000 pixels hologram (top), and the corresponding reconstruction (bottom). (d) 1000 × 1000 pixel extrapolated hologram from (c) after 300 iterations (top) and its reconstruction (bottom). Adapted from [156].

Figure 26 shows an example of extrapolation of an optical in-line hologram, which demonstrates that even if some part of the hologram is missing, it can be recovered during the iterative procedure. The sample contains four circles (Fig. 26(a)), which can be recognised in the reconstruction of the hologram (Fig. 26(b)). When only a fraction of the hologram is available (Fig. 26(c)), the reconstructed object barely resembles the original object, but when the iterative procedure with extrapolation is used, the missing part of the hologram can be restored, and the reconstructed object almost matches the original distribution (Fig. 26(d)). This leads to the conclusion that even a fraction of a hologram can be sufficient to recover the object distribution, as stated by Gabor: "This interference pattern I called a 'hologram', from the Greek word 'holos', the whole, because it contained the whole information" [160]. The extrapolation of diffraction patterns works only after an initial reconstruction, and thus an initial phase guess for the diffraction pattern is made. An example of the extrapolation method applied to the X-ray diffraction pattern is shown in Fig. 27.

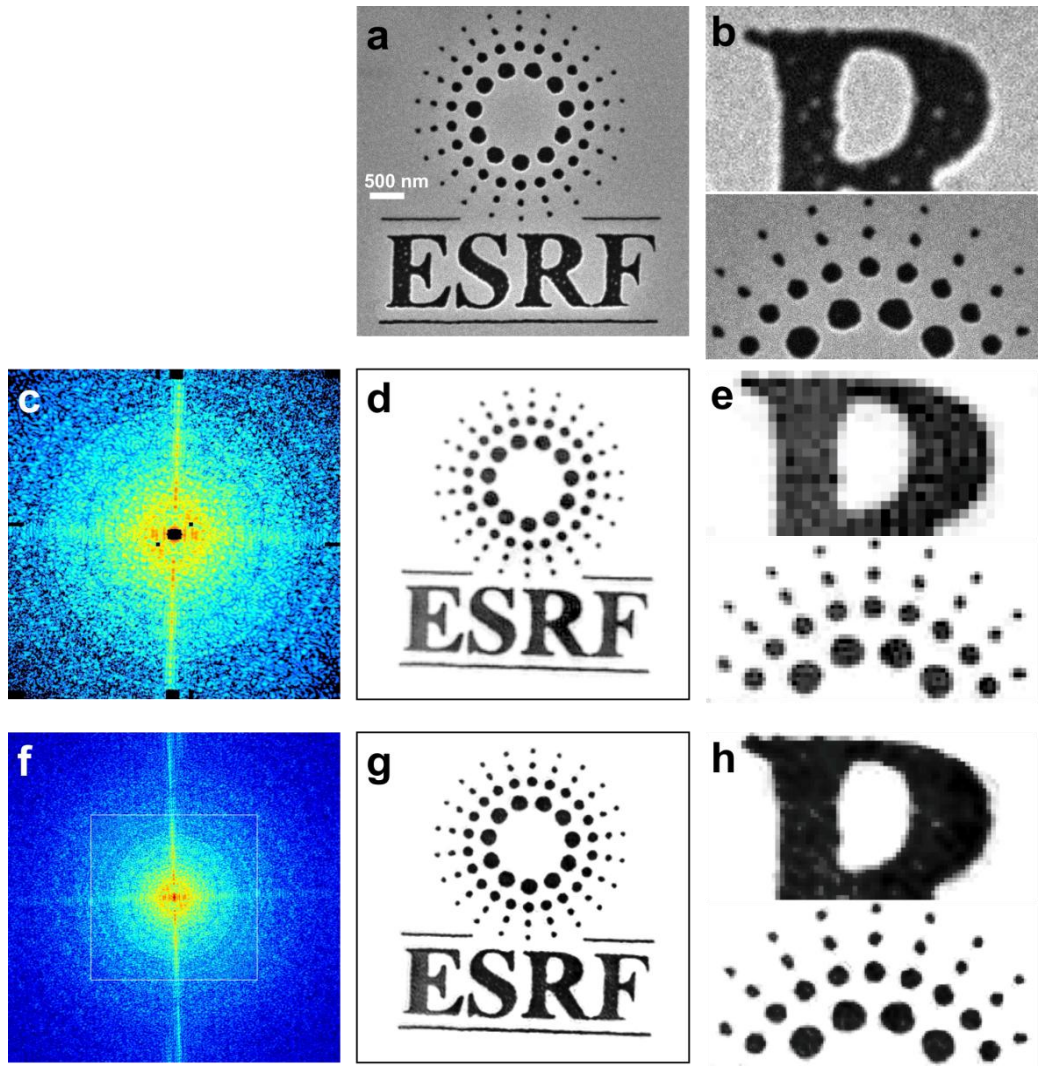

Fig. 27. Extrapolation and reconstruction of an experimental X-ray diffraction pattern. (a) Scanning electron microscopy image of the sample and (b) its magnified regions. (c) Experimental diffraction pattern, (d) sample distribution reconstructed from the diffraction pattern shown in (c) and (e) its magnified regions. (f) Diffraction pattern extrapolated from the experimental record within the white square, (g) its reconstruction, and (h) magnified regions. Reprinted from [158], with the permission of AIP Publishing.

The reason why extrapolation works can be explained as follows. Any finite signal has an unlimited spectrum and vice versa. Since we are considering an object which has a finite size, its Fourier spectrum, or the distribution of the diffracted wave in the far-field, is infinite. This is easy to image for a point-like source: the distribution of the wave originating from a point-like source is infinite in the far field. An object can be represented as consisting of such point-like scatterers, each creating a spherical wave, in agreement with the Huygens principle. A detector has a finite size, and can only capture a fraction of the distribution of the diffracted wave. However, even this captured fraction will contain information about the complete distribution of all the waves from the scatterers, and thus the complete object distribution can be reconstructed. On the other hand, in an iterative phase retrieval algorithm, the constraint that the distribution of the amplitudes in the detector plane must be equal to the measured amplitudes limited by the detector size automatically involves the constraint that these amplitudes only exist within the detector area and should be zero elsewhere. This is an unnatural constraint that does not describe any wave behaviour. Hence, extrapolation routines removes this constraint, allowing for natural wave behaviour of the recovered wavefront.

The extrapolation method is applicable to any type of interference pattern, including holograms [35, 129, 156, 161], diffraction patterns [157-159, 162], etc. Extrapolation can also be implemented into a ptychography reconstruction routine [99]. A limited size (and thus low-resolution) interference pattern is sufficient to recreate a high-resolution reconstruction of the object. This implies that even with no additional experiments, the resolution can subsequently be enhanced by applying this extrapolation technique even to previously reconstructed experimental data. This extrapolation method has drawn interest from the imaging community, as the problems of limited detector size and signal loss due to noise and associated limited resolution arise in many imaging techniques, regardless of whether imaging is carried out with electrons, X-rays, terahertz waves or any other radiation. When applied to resolution enhancement in terahertz holograms [35, 129, 161], iterative phase retrieval with extrapolation improved the resolution so that features of sub-wavelength size (35 $\mu$m) were resolved from a 2.52 THz hologram (118.83 $\mu$m wavelength) [35]. Extrapolation can therefore improve the resolution severalfold, and even allows features smaller than the wavelength to be resolved.

## 4.4 Outlook

All of the current coherent diffractive imaging techniques mentioned here, and their advanced forms such as ptychography, holographic tomography and 3D coherent diffractive imaging, are heavily reliant on mathematical methods and numerical analysis. Advanced algorithms are therefore very valuable to the coherent imaging community. Even for classical CDI, there is an ongoing search for better constraints that allow for faster and reliable reconstruction, as hundreds of successful iterative runs currently need to be averaged to give the reconstructed object distribution. In some cases, mathematical methods still to be developed. For example, all existing phase retrieval methods are based on the idea of reconstructing the object wave based on its repeated back and forth propagation between *two planes*: the detector and the object plane. As a result, only a 2D projection of a truly 3D object is recovered in the object plane. Although the relative phase shifts between the scatterers, and thus 3D information about the object, is already stored in the diffraction pattern [163], novel "outside-the-box" reconstruction methods must be invented before this information can be extracted. Currently, 3D information is extracted by performing tomographic CDI [55, 64, 164], or by the more recent advanced technique of tomographic ptychography [67, 105, 107-110, 114].

Low-energy electron in-line holography has already been demonstrated in the imaging of individual biological molecules at a resolution of about 1 nm, meaning that this technique can complement other high-resolution imaging techniques such as cryo-electron microscopy or X-ray crystallography. Unlike small-angle X-ray scattering, which also can be used to obtain the shapes of macromolecules, low-energy electron in-line holography does not require averaging over many molecules. In addition, low-energy electrons have a much higher sensitivity to local potentials than high-energy electrons, meaning that this is a perfect form of radiation for the study of 2D materials such as graphene and van der Waals structures [148-151, 165]. Moreover, low-energy electron in-line holography arrangement can be employed for mapping unoccupied electronic states of 2D materials [150].

The method of 3D volumetric deconvolution in holography [152] has already been successfully applied to 3D particle field reconstruction and particle tracking [153, 154], and for the 3D reconstruction of continuous objects [152, 155]. However, the most interesting results are expected from the application of this method to X-ray and electron experimental data, where individual atoms can be resolved by applying the 3D volumetric deconvolution method.

The extrapolation method has already generated interest in the coherent imaging community, since it can help to overcome the size limit of the detector, or can help in recovering information buried in noise. The extrapolation routine can be optimised further to deal faster and more efficiently with 3D datasets. The post-extrapolation of an interference pattern beyond experimental data, which results in enhanced resolution, has the potential to be applied to any image recorded with coherent waves, with no need for a new experiment. This would provide a fundamentally novel principle for increasing the resolution beyond the Abbe limit.